\documentclass[sn-basic]{sn-jnl}

\usepackage{graphicx}%
\usepackage{amsmath,amssymb,amsfonts}%
\usepackage{manyfoot}%

\usepackage{adjustbox}
\usepackage{multirow}%
\usepackage{xcolor}%
\usepackage{booktabs}%
\usepackage{makecell}
\usepackage{caption}
\usepackage{siunitx}
\usepackage{acronym}
\usepackage{moresize}

\usepackage{chngpage}

\usepackage{colortbl}
\definecolor{Gray}{rgb}{0.9,0.9,0.9}
\newcommand{\cc}{\cellcolor{Gray}}

\graphicspath{{Figures/}}

\begin{document}

\title[Article Title]{Uncertainties of Satellite-based Essential Climate Variables from Deep Learning}


\author*[1]{\fnm{Junyang} \sur{Gou}}\email{jungou@ethz.ch}

\author[2]{\fnm{Arnt-Børre} \sur{Salberg}}\email{salberg@nr.no}

\author[1]{\fnm{Mostafa} \sur{Kiani Shahvandi}}\email{mkiani@ethz.ch}

\author[3]{\fnm{Mohammad J.} \sur{Tourian}}\email{tourian@gis.uni-stuttgart.de}

\author[4]{\fnm{Ulrich} \sur{Meyer}}\email{ulrich.meyer@unibe.ch}

\author[5]{\fnm{Eva} \sur{Boergens}}\email{eva.boergens@gfz-potsdam.de}

\author[2]{\fnm{Anders U.} \sur{Waldeland}}\email{andersuw@nr.no}

\author[6,7]{\fnm{Isabella} \sur{Velicogna}}\email{isabella@uci.edu}

\author[2]{\fnm{Fredrik} \sur{Dahl}}\email{fadahl@nr.no}

\author[4]{\fnm{Adrian} \sur{Jäggi}}\email{adrian.jaeggi@unibe.ch}

\author[1]{\fnm{Konrad} \sur{Schindler}}\email{schindler@ethz.ch}

\author[1]{\fnm{Benedikt} \sur{Soja}}\email{soja@ethz.ch}

\affil*[1]{\orgdiv{Institute of Geodesy and Photogrammetry}, \orgname{ETH Zurich}}

\affil[2]{\orgdiv{Norwegian Computing Center}}

\affil[3]{\orgdiv{Institute of Geodesy}, \orgname{University of Stuttgart}}

\affil[4]{\orgdiv{Astronomical Institute}, \orgname{University of Bern}}

\affil[5]{\orgdiv{Department 1: Geodesy, Section 1.3: Earth System Modelling}, \orgname{GFZ Helmholtz Centre for Geosciences}}

\affil[6]{\orgdiv{Department of Earth System Science}, \orgname{University of California Irvine}}

\affil[7]{\orgdiv{Jet Propulsion Laboratory}, \orgname{California Institute of Technology}}


\abstract{Accurate uncertainty information associated with essential climate variables (ECVs) is crucial for reliable climate modeling and understanding the spatiotemporal evolution of the Earth system. In recent years, geoscience and climate scientists have benefited from rapid progress in deep learning to advance the estimation of ECV products with improved accuracy. However, the quantification of uncertainties associated with the output of such deep learning models has yet to be thoroughly adopted. This survey explores the types of uncertainties associated with ECVs estimated from deep learning and the techniques to quantify them. The focus is on highlighting the importance of quantifying uncertainties inherent in ECV estimates, considering the dynamic and multifaceted nature of climate data. The survey starts by clarifying the definition of aleatoric and epistemic uncertainties and their roles in a typical satellite observation processing workflow, followed by bridging the gap between conventional statistical and deep learning views on uncertainties. Then, we comprehensively review the existing techniques for quantifying uncertainties associated with deep learning algorithms, focusing on their application in ECV studies. The specific need for modification to fit the requirements from both the Earth observation side and the deep learning side in such interdisciplinary tasks is discussed. Finally, we demonstrate our findings with two ECV examples, snow cover and terrestrial water storage, and provide our perspectives for future research.
}


\maketitle
\section*{Highlights}
\begin{itemize}
    \item Comprehensive review of uncertainty quantification approaches using deep learning applied to earth observation data and essential climate variables (ECVs).
    \item Discuss and bridge the gap between deep learning and conventional statistical perspectives on uncertainties.
    \item Demonstrate the efficiency of deep learning uncertainty quantification approaches with example use cases for two selected ECVs: snow cover and terrestrial water storage changes.
    \item Provide recommendations for geoscience and climate scientists to accurately quantify uncertainties when applying deep learning techniques.
\end{itemize}

\section{Introduction}
\label{sec:Introduction}
The Earth's climate system has been evolving rapidly in recent years, leading to more frequent climate extremes~\citep{aghakouchak2020ClimateExtremes,rodell2023Hydroclimatic} and pressing needs for systematic observations of critical climate variables. To establish a sustainable climate monitoring system, the Global Climate Observing System (GCOS) program developed the concept of “Essential Climate Variables” (ECVs; \citealt{zemp2022GCOS}), which provide fundamental observational requirements to document and understand climate change, constrain climate prediction, and inform policy decisions on climate adaptation and mitigation~\citep{bojinski2014ECV}. ECVs are monitored using satellite systems, ground-based, ocean-based, and atmospheric sensors and form the basis for defining science requirements of observational programs of the Earth system~\citep{bayat2021toward}. Unfortunately, not all ECVs can be observed cost-effectively, and some are not even directly observable~\citep{bojinski2014ECV}. Moreover, the current observing systems for most ECVs cannot fulfill the defined ideal or intermediate requirements in terms of spatio-temporal resolution and/or accuracy. The observations may be sporadic, not acquired frequently enough, of limited accuracy, or even entirely missing over parts of the Earth system. For climate studies, the measurements need to be conducted over long time scales (decades to multiple decades), continuously, comprehensively, systematically, and globally. To this end, satellite Earth observation (sEO) data have played a critical role and have truly improved the availability and quality of observations of ECVs~\citep{anderson2017EO4SD}.

Due to the complex nature of the Earth system and our inevitably incomplete knowledge of it, data-driven methods play an important role in estimating ECVs~\citep[e.g.,][]{wylie2007Datadriven_Carbon,dangendorf2021DataDriven_SeaLevel,gou2022RiwiSAR}. Such methods become more promising with the recent progress in deep learning, which have significant potential for improving the quality of observations and models of the Earth's climate system~\citep{reichstein2019deeplearning4geoscience,schneider2023AI4ClimateModel}, especially when applied to data from sEO techniques, which have been providing a vast amount of observations of the Earth~\citep{zhu2017DLReview}. Deep learning approaches have achieved remarkable success in various fields of satellite-based climate monitoring, including pattern recognition, spatio-temporal downscaling, and quantitative analysis~\citep{yuan2020DLinRSE}. Their value and further potential have been widely recognized by geoscience and climate scientists for ECV retrievals. Many studies successfully applied deep learning techniques to various ECVs in various domains of Earth science, often with unprecedented efficiency and accuracy. However, uncertainty quantification is still missing in many studies~\citep[see, e.g., ][and Table~\ref{table:Overview_ECVs}]{jeppesen2019cloud_noUQ,segal2020cloud_noUQ,liu2020CNN4LandCover_noUQ,li2022Precipitation_noUQ}, which has been defined as a critical next step by some pioneering studies~\citep{lam2023Weather_Science_noUQ,zhang2023AI4Weather_Nature,price2023Gencast}. 

Realistic uncertainty information is crucial for interpreting or predicting changes in the climate system and, in turn, provides confidence levels for forecasts used by decision makers~\citep{reilly2001UncertaintyImportance,smith2011UncertaintyImportance}. Such uncertainties are necessary for most, if not all, further investigations based on ECV datasets, such as assimilating data into models~\citep[e.g., ][]{Schumacher2016,DeLannoy2019,Girotto2020,Luo2023,gerdener2023GLWS2}, data fusion~\citep[e.g., ][]{tourian2023Downscale}, or extreme event and time series analysis~\citep{Hoffmann2020,Laimighofer2022,saemian2024PSDI}. Multiple ECVs may be combined to derive the target parameter of interest, such as the groundwater storage changes obtained by combining multiple other satellite-based ECVs~\citep{Guentner2024}. In this case, the uncertainties of the different input data sets and their harmonization are fundamentally important. Unsatisfactory uncertainty estimates pose a significant challenge in using currently available ECV products~\citep{Soltani2021}.

However, applying appropriate deep learning approaches to quantify realistic uncertainty information in estimating ECVs based on real-world sEO data is challenging. \cite{gawlikowski2023UQinDLsurvey} categorize the uncertainties into three sources: data acquisition, deep learning model design and training, and inference. Most of the efforts so far have been dedicated to quantifying the uncertainties related to model architecture and training~\citep{abdar2021UQinDLsurvey}, but have largely ignored (or only implicitly considered) the uncertainties related to data acquisition or during inference~\citep{he2023UQinDLSource}, since the latter are not dominating in classical vision or language modeling tasks that motivated the recent revolution of deep learning techniques. However, uncertainties originating from data acquisition or inference processes are crucial for satellite data and are the keys to providing realistic uncertainty information for sEO-derived ECVs. The high variability of the real world needs to be taken into account. Otherwise, it will introduce significant distribution shifts in the data~\citep{gawlikowski2023UQinDLsurvey}. Nevertheless, EO satellites usually have multiple measurement systems with assorted characteristics onboard, the inherent noise of which needs to be characterized~\citep{chuvieco2020RS}. Another major barrier is the lack of ground truth, prohibiting us from performing uncertainty calibration/validation based on the information contained in a validation dataset~\citep{kuleshov2018CalibrationRegression}. This very problem poses a great challenge in validating satellite-based ECVs~\citep{bayat2021toward}. Furthermore, it is questionable whether the purely statistical deep learning paradigm is sufficient to quantify uncertainties related to physical domain knowledge, especially the unknown part of the total uncertainty~\citep{povey2015KnownAndUnknowns}. In summary, we need to pay attention to the special requirements for the task at hand and apply the appropriate deep learning algorithms to accurately quantify the uncertainties~\citep{rolf2024mission}.

In this paper, we comprehensively review the uncertainty quantification approaches that have been applied within deep learning frameworks to address problems related to ECVs. In Section~\ref{sec:ECV-and-Uncertainty}, we start by introducing the sources of uncertainties from a theoretical perspective and combine them with general processing pipelines of satellite observations to outline the nature of the problem. Then, we discuss the different perspectives from conventional statistics and deep learning. The bridge between the two perspectives is built based on analyzing the similarities and differences between them. Section~\ref{sec:UQ-DL} provides a review of deep learning uncertainty quantification approaches with a focus on those that have been applied in estimating ECVs. Section~\ref{sec:Applications} presents two use cases by applying various deep learning algorithms to estimate two selected ECVs: snow cover fraction (SCF) and terrestrial water storage (TWS), discussing the different characteristics of derived uncertainties. Finally, we summarize our discussions and provide an outlook in Section~\ref{sec:Conclusions and Future Perspectives}.

\section{ECV from Satellite Observations and Their Uncertainties}
\label{sec:ECV-and-Uncertainty}
\subsection{Overview of ECVs from Satellite Observations}
GCOS currently specifies 55 ECVs, which are separated into three main domains: terrestrial, ocean, and atmosphere, and further sub-categorized into ten subdomains. In all the domains, satellite observations play a crucial role by providing global coverage of targeted variables with consistent quality, resulting in substantial contributions to determining ECVs. Previous studies have reported that satellite Earth observation data contribute to 33 ECVs significantly~\citep{miranda2020reviewingECVs} and 42 ECVs at least partially~\citep{giuliani2020ECVs,ballari2023ECVsFromEO}. Recent developments have increased this number to 51 out of 55 ECVs, which are at least partially estimated from sEO data.  Space agencies worldwide have supported ambitious programs to develop measurement approaches, algorithms for retrieving ECVs from satellite observations, output products that document ECVs, and maintain continuous services to the community. Some exemplary platforms are provided by the European governments, including the Copernicus Climate Change Service (C3S; \citealt{buontempo2022C3S}); the ESA Climate Change Initiative (ESA CCI; \citealt{plummer2017ESACCI}); the Copernicus Land Monitoring Service (CLMS); EUMETSAT, and the US government, including the Physical Oceanography Distributed Active Archive Center (PODAAC), the National Oceanic and Atmospheric Administration (NOAA), and the NASA MEaSUREs program. These services have made ECV products more easily accessible globally to the broader community and have helped document progress toward global sustainable development targets~\citep{anderson2017EO4SD}, improve the quality of reanalysis products, improve climate models through data assimilation approaches, and also help support critical decisions and policies related to the management of our resources, e.g. groundwater resources~\citep{springer2023Space-based,tapley2019GRACE4Climate}.

The geosciences and climate scientists have widely realized the potential of deep learning algorithms, which can be applied to most, if not all, of the ECVs. A summary of these studies is given in Table~\ref{table:Overview_ECVs}. A common idea is to derive the relationship between sEO-based ECVs and in-situ measurements using deep learning algorithms to improve the accuracy and spatio-temporal resolutions of the sEO-based ECV products~\citep[e.g.,][]{Yang_2021_soilcarbon,Kolluru_2022_oceancolor,Koppa_2022_evaporation,Cui_2023_snow,Chen_2024_permafrost}. We can also benefit from this strategy to build empirical relationships between ECVs that are not directly observable from satellite measurements and other sEO-derived quantities to enable a quasi-sEO estimation~\citep{Hamrani_2020_greenhouse}. Clarifying the possible uncertainty sources in sEO data processing pipelines and linking them to the deep learning uncertainty quantification approaches is crucial for delivering realistic uncertainty information. Moreover, most of the applications of deep learning to ECV estimations deal with regression problems (i.e., providing continuous numerical value) with some exceptions, such as land cover classification~\citep{Jagannathan_2021_landcover} or fire detection~\citep{Sathishkumar_2023_fire} requiring classification or segmentation (also known as pixel-wise classification). Given this trend, we will focus on uncertainty quantification approaches for regression problems using deep learning techniques and try to bridge them with the classical statistical methods in the remainder of this paper.

\begin{table}[!htbp]
    \caption{Overview of ECVs. The possible contributions from satellite observations and types of deep learning methods are listed. The ECVs that cannot be directly observed from satellite missions are marked in grey. The potential machine learning algorithms are categorized into regression (R) and classification (C). We note that the regression and classification tasks in this table also include grid-wise regression and classification, which are usually understood as segmentation tasks. Special notions:	$^\ast$ denotes the studies with uncertainty quantification approaches; $^\dag$ denotes the studies without using sEO data.}
    \label{table:Overview_ECVs}

    \ssmall
    \centerline

    \begin{tabular}{ccccccc}
        \toprule
        Domain& Subdomain& Variables& \makecell{Type of tasks}& \makecell{References}& \makecell{Exemplary \\data provision}\\
        \midrule
        \multirow{20}{*}{Terrestrial}& \multirow{6}{*}{Hydrosphere}& Groundwater& R& \cite{miro2018downscalingGroundwater}& C3S\\
                                     &                             & Lakes& R \& C& \cite{mullen2023using}& C3S\\
                                     &                             & River discharge&   R& \cite{ansari2023RivQNet}$^\dag$& ESA CCI\\
                                     &                             & Soil moisture& R \& C  & \cite{Singh_2023_soilmoisture}$^*$& C3S\\
                                     &                             & Evaporation from land&  R & \cite{Koppa_2022_evaporation}& OpenET\\
                                     &                             & Terrestrial water storage& R& \cite{gou2024global}$^*$& C3S\\\cmidrule{2-6}
                                     & \multirow{4}{*}{Cryosphere}& Glaciers&  R \& C& \cite{Thomas_2023_glaciers}& C3S\\
                                     &                            & Ice sheets and ice shelves&   R \& C& \cite{Tollenaar_2024_icesheet}$^*$& C3S\\
                                     &                            & Permafrost&   R&  \cite{Chen_2024_permafrost}& ESA CCI\\
                                     &                            & Snow&   R \& C& \cite{Cui_2023_snow}& CLMS\\\cmidrule{2-6}
                                     & \multirow{8}{*}{Biosphere}& Above-ground biomass&  R & \cite{Zhang_2019_biomass}& ESA CCI\\
                                     &                            & Albedo&  R & \cite{Chen_2023_albedo}& C3S\\
                                     &                            & Fire&  R \& C & \cite{Sathishkumar_2023_fire}$^\dag$& ESA CCI\\
                                     &                            & FAPAR\footnote{Fraction of Absorbed Photosynthetically Active Radiation}&  R & \cite{Ma_2022_fapar}& ESA CCI\\
                                     &                            & Land cover& C & \cite{Jagannathan_2021_landcover}& CLMS\\
                                     &                            & Land surface temperature&  R & \cite{Wang_2021_landsurfacetemperature}& CLMS\\
                                     &                            & Leaf area index&  R& \cite{Castro-Valdecantos_2022_leafareaindex}& ESA CCI\\
                                     &                            & Soil carbon&  R & \cite{Yang_2021_soilcarbon}& - \\\cmidrule{2-6}
                                     & \multirow{2}{*}{Anthroposphere}& \cc Anthropogenic Greenhouse gas fluxes& R \cc & \cc \cite{Hamrani_2020_greenhouse}& \cc -\\
                                     &                                & \cc Anthropogenic water use& \cc R & \cc \cite{Wunsch_2022_wateruse}$^{*\dag}$& \cc -\\
        \midrule
        \multirow{19}{*}{Ocean}& \multirow{11}{*}{Physical}& Ocean surface heat flux & R & \cite{George_2021_heatflux}& NOAA\\
                               &                           & Sea ice & R \& C& \cite{Andersson_2021_seaice}$^*$& C3S\\
                               &                           & Sea level & R & \cite{Nieves_2021_sealevel}$^*$& C3S\\
                               &                           & Sea state & R & \cite{Mittendorf_2022_seastate}$^\dag$& CMS\\
                               &                           & Sea surface currents & R & \cite{Sinha_2021_seacurrents}$^\dag$& CMS\\
                               &                           & Sea surface salinity & R & \cite{Guillou_2023_seasalinity}$^\dag$& CMS\\
                               &                           & Sea surface stress & R & \cite{Yousefi_2024_seastress}$^\dag$& -\\
                               &                           & Sea surface temperature & R & \cite{Xiao_2019_seatemperature}$^*$& C3S\\
                               &                           & Subsurface currents & R & \cite{Bradbury_2021_subsurfacecurrents}$^\dag$& CMS\\
                               &                           & Subsurface salinity & R & \cite{Tian_2022_subsurfacesalinity}$^*$& CMS\\
                               &                           & Subsurface temperature & R & \cite{Su_2022_subsurfacetemperature}& CMS\\\cmidrule{2-6}
                               & \multirow{6}{*}{Biogeochemical}& Inorganic carbon& R & \cite{Galdies_2023_inorganiccarbon}& CMS\\
                               &                                & Nitrous oxide& R& \cite{Yang_2020_nitrousoxide}$^{*\dag}$& CMS\\
                               &                                & Nutrients& R & \cite{Contractor_2021_nutrients}$^{*\dag}$& CMS\\
                               &                                & Ocean colour& S & \cite{Kolluru_2022_oceancolor}& C3S\\
                               &                                & Oxygen& R & \cite{Sharp_2023_oceanoxygen}$^{*\dag}$& CMS\\
                               &                                & \cc Transient tracers& \cc R & \cc -& \cc -\\\cmidrule{2-6}
                               & \multirow{2}{*}{\makecell{Biological/\\ecosystems}}& Marine habitats& R \& C & \cite{Rubbens_2023_marinehabitats}& -\\
                               &                                       & Plankton& R & \cite{Ciranni_2024_plankton}$^\dag$& CMS\\
        \midrule
        \multirow{16}{*}{Atmosphere}& \multirow{6}{*}{Surface}& Precipitation& R & \cite{Gavahi_2023_precipitation}& C3S\\
                                    &                         & \cc Pressure& \cc R & \cc \cite{Karmakar_2023_pressure}$^\dag$& \cc -\\
                                    &                         & Radiation budget& R & \cite{Li_2022_atmosphericradiation}& C3S\\
                                    &                         & Temperature&  R & \cite{Shen_2020_atmospherictemperature}& EUMETSAT\\
                                    &                         & Water vapour& R & \cite{Wu_2023_atmosphericwatervapour}& C3S\\
                                    &                         & Wind speed and direction& R & \cite{Jiang_2024_atmosphericwind}$^\dag$& EUMETSAT\\\cmidrule{2-6}
                                    & \multirow{6}{*}{Upper-air}& Earth radiation budget& R & \cite{Yao_2023_atmosphericradiation}$^\dag$& C3S\\
                                    &                           & Lightning& R & \cite{Zhou_2020_atmosphericlightning}$^*$& NASA\\
                                    &                           & Temperature & R & \cite{Haynes_2024_upperatmospherictemperature}$^*$& EUMETSAT\\
                                    &                           & Water vapour & R & \cite{zhang2024PWV}& NASA\\ 
                                    &                           & Wind speed and direction & R & \cite{Das_2021_atmosphericupperairwindwpeed}$^\dag$& EUMETSAT\\
                                    &                           & Clouds& R \& C & \cite{Wright_2024_atmosphericclouds}$^*$& C3S\\\cmidrule{2-6}
                                    & \multirow{4}{*}{\makecell{Atmospheric \\Composition}}& Aerosols& R & \cite{Tao_2023_atmosphericaerosols}& ESA CCI\\
                                    &                                         & Greenhouse gases& R& \cite{Altikat_2021_greenhouse}$^\dag$& -\\
                                    &                                         & Ozone& R & \cite{Han_2023_atmosphericozone}& -\\
                                    &                                         & Precursors for aerosols and ozone& R & \cite{Tao_2024_atmosphericprecursors}& -\\
        \bottomrule
    \end{tabular}
\end{table}

\subsection{Different uncertainties within satellite-based estimating pipeline}
\subsubsection{Definition of uncertainties and their coupled nature}
Uncertainties can be categorized into \textit{aleatoric uncertainty} (data uncertainty) and \textit{epistemic uncertainty} (model uncertainty). Aleatoric uncertainty is caused by the inherent randomness of an event, such as uncertainties contained in forcing data or unavoidable measurement errors. Therefore, this type of uncertainty is considered irreducible. On the contrary, epistemic uncertainty is caused by deficiencies in the model, also known as lack of knowledge. We note that the reasons for model deficiencies in a conventional statistical model can be fundamentally different from the ones for a deep learning model, which will be discussed later in Section~\ref{sec:Conventional_Statistical_and_Deep_Learning}. Here we try to clarify the source of aleatoric and epistemic uncertainties in a deep learning model based on the law of total variance. In a deep learning model, the trainable parameters $\boldsymbol{\theta}$ are usually estimated based on a given training dataset $\mathcal{D}=\{\mathbf{X},\mathbf{Y}\}$. The inputs $\mathbf{X}=[\mathbf{x}_1,\dotsc,\mathbf{x}_n]^\mathrm{T}\in\mathbb{R}^{n\times d}$ contains $n$ vectors with $d$ features, whereas the targets $\mathbf{Y}=[y_1,\dotsc,y_n]^\mathrm{T}\in\mathbb{R}^{n\times1}$ contains $n$ values~\citep[may also generalize to multi-dimensional vectors, see, e.g.,][]{goodfellow2016DL}. Once we apply this model to a new and unseen input $\mathbf{x}^*$ (inference), we can obtain the corresponding prediction $y^*$ with the associated total variance as:
\begin{equation}
    \mathrm{Var}\left(y^*|\mathbf{x}^*\right) = \underbrace{\mathbb{E}_\theta\left[\mathrm{Var}_{y^*}\left(y^*|\mathbf{x}^*,\boldsymbol{\theta}\right)\right]}_\mathrm{Aleatoric} + \underbrace{\mathrm{Var}_{\theta}\left(\mathbb{E}_{y^*}\left[y^*|\mathbf{x}^*,\boldsymbol{\theta}\right]\right)}_\mathrm{Epistemic}, \label{eq:Total variance}
\end{equation}
where the first term $\mathbb{E}_\theta[*]$ describes the variability of the prediction $y^*$ averaged over all the possible model parameters, quantifying the uncertainties caused by the inherent randomness of data regardless of the model parameters. Hence, it provides a rigorous definition of aleatoric uncertainties. The second term $\mathrm{Var}_{\theta}(*)$ measures the uncertainties of the mean of $y^*$ given each model realization $\boldsymbol{\theta}$. By computing the expectation of all possible $y^*$, the inherent data randomness is averaged out, and therefore, the second term is understood as epistemic uncertainty~\citep{krause2022PAI}. However, we should note that the definition given in Eq.~\eqref{eq:Total variance} refers to uncertainty categories during inference based on the parameters $\boldsymbol{\theta}$ following the distribution of $p\left(\boldsymbol{\theta}|\mathcal{D}\right)$ with the given training dataset $\mathcal{D}$. The defined epistemic uncertainty depends on the data (aleatoric) uncertainty of the training dataset (both inputs and targets), revealing the mixed nature of aleatoric and epistemic uncertainties. It has been argued that the aleatoric and epistemic uncertainties cannot be unambiguously separated~\citep{hora1996aleatory}, in particular for real-world data \citep{kahl2024}. Recently, \cite{gruber2023SourcesOfUncertainty} have given a comprehensive overview of the source of uncertainties in machine learning from a statistical view and confirmed this issue. With these thoughts, we clarify the definition of aleatoric and epistemic uncertainties used in this paper as follows:
\begin{itemize}
    \item \textbf{Aleatoric uncertainties}: Uncertainties caused by data problems, including training and inference data. This category includes irreducible uncertainty due to inherent noise in the observational systems and outliers caused by problematic measurements or observing environments. Thus, we can confidently refer to them as data uncertainties.
    \item \textbf{Epistemic uncertainties}: Uncertainties caused by the imperfect model. The possible reasons include inadequate assumptions,difficulties in optimization (local optimum), inadequate training (under- or over-fitting), insufficient model capacity, etc.
\end{itemize}

Although the aleatoric and epistemic uncertainties cannot be precisely separated, the total uncertainties are defined in Eq.~\eqref{eq:Total variance} with values independent of different categorizing strategies. Therefore, our ultimate goal is to quantify the total uncertainties as realistically as possible without overemphasizing the distinction between different types of uncertainties. As a matter of fact, the total uncertainties are the most crucial information for decision-makers. 

\subsubsection{Data processing pipeline and sources of uncertainties}
\label{sec:Data processing pipeline and sources of uncertainties}
\begin{figure}[!ht]
    \centering
    \includegraphics[width=11.9cm]{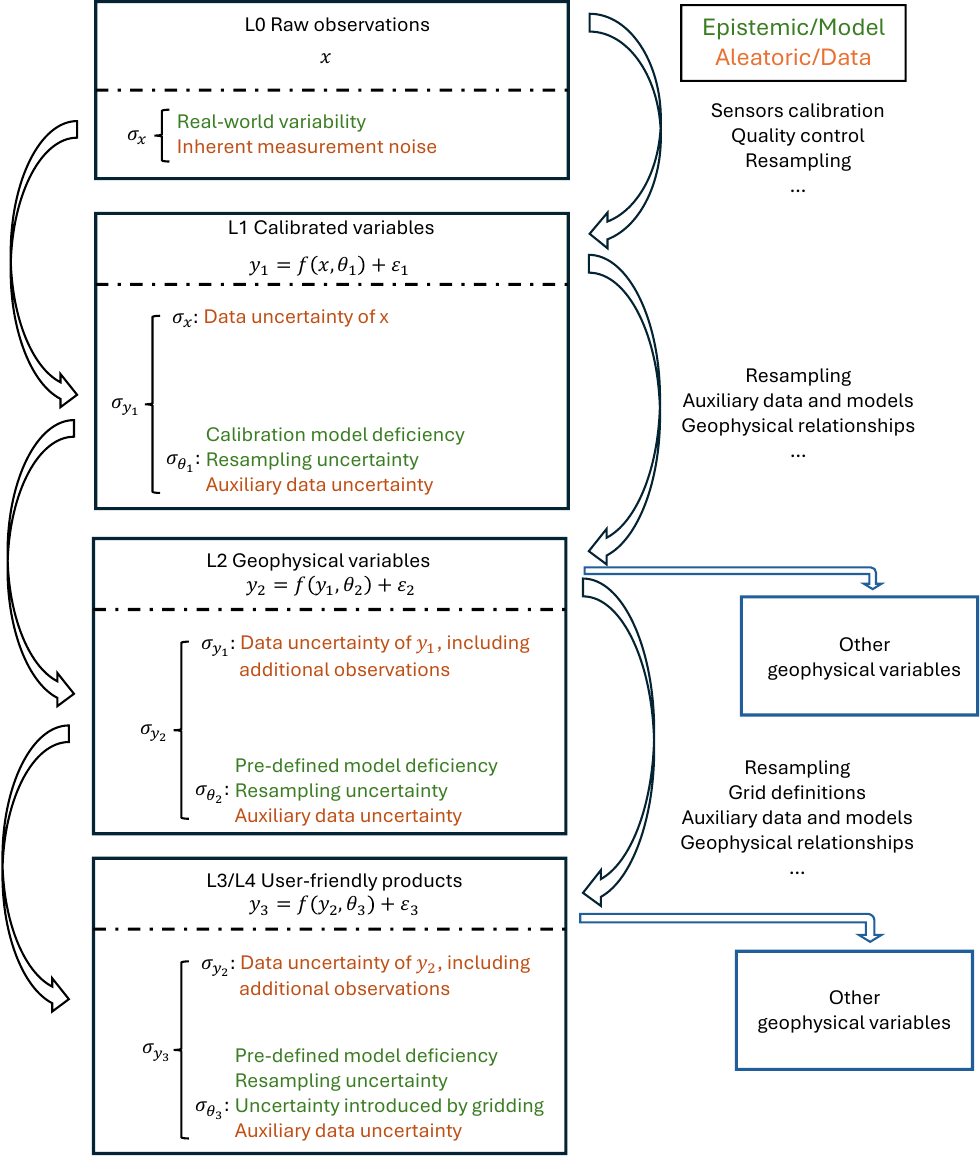}
    \caption{Estimating pipeline of sEO-based variables from L0 to L4 and the corresponding uncertainty propagation. The uncertainties from models (epistemic) are denoted in green, whereas those from data (aleatoric) are denoted in orange. The right arrows denote the processing steps with necessary actions or additional information, whereas the left arrows indicate uncertainty propagation. The deep learning algorithms are usually applied to L2 to L4 products to derive other geophysical variables depicted in blue.}
    \label{fig:Estimating_pipeline}
\end{figure}

To better understand the sources of uncertainties in a typical sEO-based ECV processing pipeline, we briefly describe the multiple levels of products and their associated uncertainty propagation~\citep{povey2015UncertaintyInRS,mittaz2019applying}. Fig.~\ref{fig:Estimating_pipeline} depicts a typical estimating pipeline with both data (right side) and uncertainty (left side) propagation. From the raw observations to user-friendly products, the satellite data are processed through multiple levels, from Level-0 (L0, indicating raw measurements) to Level-1 (processed products with telemetry), Level-2 (geocoded, calibrated products), Level-3 (gridded products over the large scale) and finally Level-4 (L4, indicating highly processed products for specific targets) using a set of algorithms documented in Algorithm Theoretical Basis Documents (ATBDs). For each individual processing step, varying uncertainty sources may be involved, as shown in Fig.~\ref{fig:Estimating_pipeline}. The primary uncertainty resources for L0 data are the real-world variability and the inherent sensor noises. The L0 data uncertainty is the primary aleatoric source for L1 products with additional contribution with other auxiliary data uncertainties, whereas the used calibration models and processing strategies introduce epistemic uncertainties. All of these uncertainties are then together propagated to L1 products and become the aleatoric uncertainties in the L1 to L2 process. At this stage, more models are involved, such as background models or predefined geophysical relationships, causing increasing impacts of epistemic uncertainties. The processing steps from L2 onward usually involve multiple data and models for processing and corrections. Therefore, it is hard to judge if aleatoric or epistemic sources dominate the total uncertainties. It is undoubtedly necessary to rigorously consider all the uncertainty sources for a realistic final uncertainty quantification. The uncertainties in individual processing steps accumulate and propagate to the final L3 or L4 products for the users to make decisions. We should note that the practical processing levels of different satellite missions may slightly differ from each other. For example, the L1 steps is sometime further separated into L1a and L1b steps and not all the generated products are avaliable to the users. An example case is the GRACE(-FO) mission which is described in Appendi~\ref{appendix:Processing pipeline of GRACE(-FO) data and associated uncertainty sources}.

Furthermore, we highlight that different satellite-based ECVs rely on varying measuring principles, and therefore, the dominating sources of data uncertainties may be different. For example, ECVs based on imaging, namely the classical remote sensing at all the different wavelengths of interest, are dominated by the sensitivity and resolution of the equipped sensors as well as the observing environments (real-world variability), such as cloudy days~\citep{toth2016ReviewRSSensor}. On the contrary, for ECVs that rely on geometry (e.g., satellite altimetry) orbit accuracy, especially in the radial direction, is of critical importance~\citep{abdalla2021ReviewAltimetry}. For ECVs based on satellite gravimetry, the requirements for accuracies of multiple sensors further increase, including the inter-satellite ranging system and accelerometers for removing non-gravitational forces~\citep{landerer2020}. The situation is similar to the potential epistemic uncertainty sources. Varying models or predefined relationships based on domain expertise exist at the level of the algorithms used to retrieve ECVs from the sEO measurements. For instance, several ECVs are deduced from the reflectance or emission of electromagnetic waves from the ocean or land surface or atmospheric layers via (more primitive) regression analysis or (more advanced) semi-empirical, physically-based retrievals, such as chlorophyll concentration from ocean color~\citep{Elachi2006, Martin2004}. There may also be inherent limitations associated with these retrieval, for instance, uncertainties associated with the depth of the water column leading to sea surface temperature or the depth of soil layers leading to the retrieval of soil moisture. For the ECVs derived from signals that are tightly coupled with other geophysical signals (e.g., satellite gravimetry), the background model to isolate the signals of interest is critical~\citep{shihora2022AOD1BRL07}. We further provide a detailed example in Appendix~\ref{appendix:Processing pipeline of GRACE(-FO) data and associated uncertainty sources} based on the whole processing pipeline of the GRACE(-FO) satellite data to clarify the complex sources of uncertainty. The exemplary pipeline again emphasizes the mixing nature of aleatoric and epistemic uncertainty. It may not be so critical to distinguish between them explicitly, but carefully considering the uncertainties in all steps is the key to reaching realistic product uncertainties and making optimal decisions.

\subsection{Conventional statistical and deep learning views on uncertainty quantification}
\label{sec:Conventional_Statistical_and_Deep_Learning}
\begin{figure}[!ht]
    \centering
    \includegraphics[width=11.9cm]{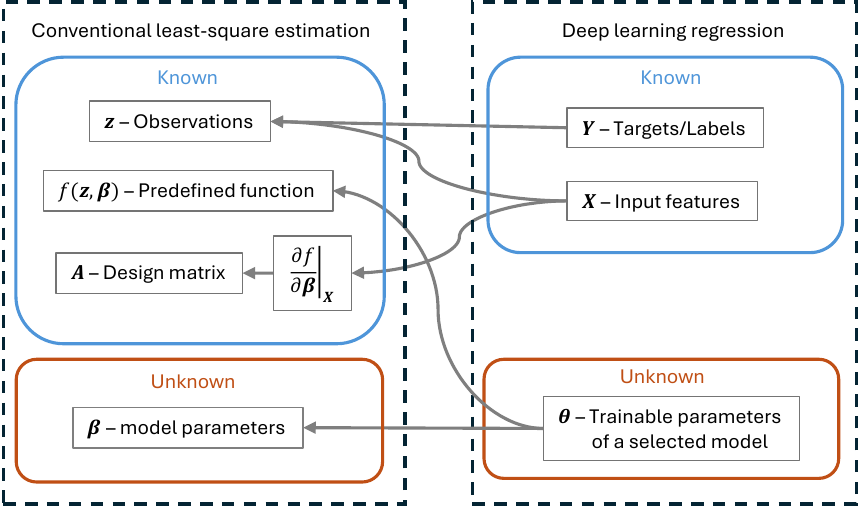}
    \caption{A schematic comparison of a conventional least-square estimation model and a deep learning regression model for the training (or parameter estimation) process. The observations in a least-square estimation problem include both inputs and targets in a deep learning model. The unknown parameters and predefined function in a classical least-square estimation model are represented by a set of unknown but trainable parameters $\boldsymbol{\theta}$ of a selected deep learning model.}
    \label{fig:Compare_LSQ_DL}
\end{figure}

This section will clarify the sources of uncertainty from conventional statistical and deep learning points of view based on theoretical considerations. A least-square estimation model and a deep learning regression model share many common settings but also have different views, which cause different sources of uncertainties and ways to quantify them. To better highlight the differences between these two views, we need to clarify a fundamental difference in estimating parameters in a conventional statistical approach and a deep learning framework. Fig.~\ref{fig:Compare_LSQ_DL} shows basic settings of a conventional least-square estimation and a deep learning regression problem for the training (or parameter estimation) process~\citep[see][for further discussion]{butt2021MLandGeodesy,kutz2023ML4PE,amiri2024DLinLSQ}.

In a typical least-square estimation, we are interested in determining a set of variables (denoted by $\boldsymbol{\beta}\in\mathbb{R}^{m\times1}$, where $m$ denotes the number of parameters) that are usually not directly measurable or their direct measurements cannot reach the expected accuracy, but can be linked to a set of other observable data based on a known (physics-based or empirical) relationship. It implies that in a least-square estimation, we typically solve an inverse problem by formulating the target variables as unknown parameters $\boldsymbol{\beta}$ in a predefined function $f\left(\mathbf{z}, \boldsymbol{\beta}\right)=0$ related to direct observations $\mathbf{z}\in\mathbb{R}^{n\times1}$, where $n$ indicates the number of samples. The function $f$ may contain both the linearized relationship (parameterized by the design matrix $\mathbf{A}$) and conditional constraints (parameterized by the constraint matrix $\mathbf{B}$). By minimizing the root mean square residuals of the predefined function, we can find the optimal estimation of target variables and obtain their uncertainties based on the law of uncertainty propagation~\citep{koch2013parameter}. In this aspect, the resulting uncertainties arise from input data uncertainties (aleatoric) under the given relationship $f$. The impacts of the model deficiency (epistemic uncertainty), usually known as the “$+0$” term, are unavoidably present but usually assumed to be small since the predefined function is assumed to be satisfactory accurate~\citep{mittaz2019applying}.

In a deep learning regression model, the functional relationship is typically unknown. Instead, the input-target pairs for the training set $\mathcal{D}=\{\mathbf{X},\mathbf{Y}\}$ are available. We choose a deep learning model with a set of trainable parameters $\boldsymbol{\theta}$ and estimate the relationship by minimizing the differences between predicted values $\mathbf{\hat{Y}}$ and labels $\mathbf{Y}$. The obtained model will be later applied to an unseen input set $\mathbf{X}^*$ to get the corresponding $\mathbf{Y}^*$ as the final predictions. Most efforts are usually put into quantifying the epistemic uncertainties, namely the uncertainties caused by $\boldsymbol{\theta}$, but the uncertainties caused by data uncertainty of both training and test sets may be overlooked~\citep{gruber2023SourcesOfUncertainty}. One of the plausible reasons is that the data uncertainty is usually inaccessible for common data types used in deep learning, such as RGB-images~\citep{deng2009imagenet}. Therefore, most of the related investigations focus on the uncertainties caused by different model architectures and optimization processes.

From Fig.~\ref{fig:Compare_LSQ_DL}, we can understand the similarities of the two approaches. Certain sets of known data are available for the training process, and some target parameters are to be estimated. The primary difference is whether or not a strong prior knowledge is considered. A least-square model converts the raw inputs into the rows of the design matrix $\mathbf{A}$ based on a strong prior knowledge imposed by the functional relationship $f$, which is designed based on domain expertise. The construction of the design matrix can be understood as a feature engineering~\citep{zheng2018FeatureEngineering} which is usually needed for a machine learning model but not highly required for a deep learning model. The recent developments in deep learning models tend to feed the raw data $\mathbf{X}$ into the model and believes the vast model capacity can exploit the information contained in data. The different views are also reflected in the target (unknown) parameters. The model parameters $\boldsymbol{\beta}$ are tightly coupled with the predefined function $f$, whereas the trainable parameters $\boldsymbol{\theta}$ are rarely constrained by prior knowledge, except for the operator mechanisms involved in the model architecture, such as the shift-invariant feature of a convolutional kernel.

The above discussion also reveals the different perspectives on uncertainty quantification and the fundamental differences in the sources of uncertainties. The epistemic uncertainty in a conventional statistical model mainly comes from the misspecification of the pre-defined model (the “$+0$” term) because the sum of unmodelled effects rarely follows a zero-mean Gaussian distribution in practice. In this case, the pre-defined relationship lacks knowledge of relevant effects because the necessary parameters to quantify them are not provided. So, the model cannot capture all the important effects or patterns in the data to reach the “best” estimation. This phenomenon can be understood as a typical underfitting issue since the conventional statistical model is typically low-dimensional (compared to deep learning) with high data redundancy. Section~\ref{sec:Statistical view} will further clarify this issue with analytical expression. On the contrary, the model capacities of deep learning models are rarely a problem since they are typically massively over-parameterized. However, a huge number of trainable parameters and complicated model architectures may cause overfitting issues and high epistemic uncertainty. These issues require careful design of the deep learning model and can be eliminated by applying advanced optimizing strategies with careful diagnoses of the loss curves~\citep{moradi2020Survey_Regularization,tian2022Survey_Regularization}. Another rather dominating reason for epistemic uncertainty in a deep learning model is the incomplete (and thus biased) training data set. If the training data set is not fully representative and cannot cover the full data distribution, the obtained model may over-rely on the general pattern shown in the training set but does not exist in the test set and discard the possible patterns that are simply not covered by the training set. The origin of the epistemic uncertainty and the ways to quantify them with ECV application examples are introduced in Section~\ref{sec:Statistical view} and Section~\ref{sec:UQ-DL}, respectively.

Regarding the aleatoric uncertainty, the conventional statistical model can rigorously consider input uncertainty and also their covariances in the frame of uncertainty propagation whenever we have a sample with associated uncertainty information. However, the uncertainty propagation approach is again highly dependent on the predefined relationship. Therefore, the aleatoric uncertainties are mixed with the epistemic uncertainties in this case (see Section~\ref{sec:Statistical view}). The deep learning techniques to quantify aleatoric uncertainty are rather limited compared to the ways to quantify epistemic uncertainty. The logic is to let the model "judge" its prediction by predicting the associated uncertainty together and formulate them in a different loss term~\citep{kendall2017WhatUncertainties}. An obvious limitation, especially for applications in satellite-based ECV studies, is the absence of treatment of input feature uncertainties, see Section~\ref{sec:Deep learning view}, and Section~\ref{sec:DEMC}.

\subsubsection{Conventional statistical view}
\label{sec:Statistical view}
The most common uncertainty quantification approach incorporated in the processing pipelines for satellite-based ECVs is based on the frequentist approach, which is grounded in long-run frequencies and relies heavily on techniques such as confidence intervals and hypothesis testing~\citep{lehmann2005testing,teunissen2006testing}. While confidence intervals offer a range of values within which the true parameter is expected to lie with a specified probability \citep{efron1979bootstrap}, hypothesis testing involves making decisions about the validity of specific hypotheses based on sample data~\citep{neyman1933testing}. Building upon these fundamentals, one widely used method in the frequentist approach for uncertainty quantification is to rely on maximum likelihood estimation (MLE). MLE seeks to find the parameter values that maximize the likelihood function, which measures the probability of the observed data given a set of parameters. The method provides estimates with desirable properties such as consistency, efficiency, and asymptotic normality~\citep{casella2002perfect}. Considering a linear relationship and assuming the variables follow Gaussian distributions, MLE is equivalent to the least squares method, which is one of the most favorable approaches for uncertainty quantification within the frequentist framework. In this context, uncertainty quantification is typically performed by deriving a covariance matrix for the unknown parameters. To illustrate this, consider the Gauss-Helmert model~\citep{teunissen2004adjustment}, which is a combination of the Gauss-Markov model (A-model) and the conditional adjustment (B-model), see \cite{koch2013parameter} for further discussion. In such a mixed model with both target parameter vectors $\boldsymbol{\beta}$ and observation vectors $\mathbf{z} $ with associated covariance matrix $D(\mathbf{z}) = \mathbf{Q}_z$ being involved, the model is typically written as:
\begin{equation}
    f(\mathbf{z},\boldsymbol{\beta}) \overset{!}{=} \mathbf{0}, \label{eq:lsq_function}
\end{equation}
meaning that the target is to minimize the misfit residuals, which is achieved by minimizing mean-square errors in a classical least-squares approach.

Eq.~\eqref{eq:lsq_function} indicates that we have knowledge of the relationships between the observations and the target parameters and can also provide knowledge about constraints applied on target parameters themselves (conditional adjustment). In reality, Eq.~\ref{eq:lsq_function} is rarely linear, so we establish Taylor points ($\boldsymbol{\beta}_0$, $\mathbf{z}_0$) for both the unknowns $\boldsymbol{\beta}$ and observations $\mathbf{z}$ as:
\begin{align}
    \widetilde{\boldsymbol{\beta}} &= \boldsymbol{\beta}_0 + \Delta\boldsymbol{\beta}, \\
    \widetilde{\mathbf{z}} &= \mathbf{z} + \mathbf{e}  \notag = \mathbf{z} - \mathbf{z}_0 + \mathbf{z}_0 + \mathbf{e} \notag = \mathbf{z}_0 + \Delta \mathbf{z}+\mathbf{e},\\
\end{align}
with $\mathbf{e}$ denoting the misfit residuals, so that the model can be linearized as:
\begin{align}
    f(\widetilde{\mathbf{z}},\widetilde{\boldsymbol{\beta}}) &= f(\mathbf{z}_0 + \Delta \mathbf{z} + \mathbf{e},\boldsymbol{\beta}_0 + \Delta \boldsymbol{\beta}) \notag \\
    &= \underbrace{f(\mathbf{z},\boldsymbol{\beta})\big|_0}_{\mathbf{w}_0} + \underbrace{\frac{\partial{f(\mathbf{z},\boldsymbol{\beta})}}{\partial{\mathbf{z}}}\bigg|_0 }_{\mathbf{B}^{\text{T}}} (\Delta \mathbf{z} + \mathbf{e}) \label{eq:Talyor}
    + \underbrace{\frac{\partial{f(\mathbf{z},\boldsymbol{\beta})}}{\partial{\boldsymbol{\beta}}}\bigg|_0 }_{\mathbf{A}} \Delta \boldsymbol{\beta} \\
    &= \underbrace{\mathbf{w}_0 + \mathbf{B}^{\text{T}}\Delta \mathbf{z}}_\mathbf{w} + \mathbf{B}^{\text{T}}\mathbf{e} + \mathbf{A}\Delta \boldsymbol{\beta} \\
    &= \mathbf{w} + \mathbf{B}^{\text{T}}\mathbf{e} + \mathbf{A}\Delta \boldsymbol{\beta} \overset{!}{=} \mathbf{0}.\label{eq:Cost_GM}
\end{align}
With such a formulation the cost function Eq.~\eqref{eq:Cost_GM} is $\mathbf{e}^{\text{T}}\mathbf{P}\mathbf{e}$ with $\mathbf{P}$ being the weight matrix for observations, defined as $\mathbf{P} = \sigma_0^2 \mathbf{Q}_z^{-1}$. The solutions take the forms:
\begin{align}
    \Delta \hat{\boldsymbol{\beta}} &= -[\mathbf{A}^{\text{T}}(\mathbf{B}^\text{T}\mathbf{P}^{-1}\mathbf{B})^{-1}\mathbf{A}]^{-1}\mathbf{A}^\text{T}(\mathbf{B}^\text{T}\mathbf{P}^{-1}\mathbf{B})^{-1}\mathbf{w}\label{eq:LSQ_weight_parameter}, \\
    \hat{\mathbf{e}} &= -\mathbf{B}\mathbf{P}^{-1}(\mathbf{B}^\text{T}\mathbf{P}^{-1}\mathbf{B})^{-1}(\mathbf{w} + \mathbf{A}\Delta\hat{\boldsymbol{\beta}}),
\end{align}
with a-posteriori estimate of $\sigma_{0}^{2}$ as

\begin{equation}
    \hat{\sigma}_{0}^{2} = \frac{\hat{\mathbf{e}}^{\text{T}}\mathbf{P}\hat{\mathbf{e}}}{\mathrm{d.o.f.}},
\end{equation}   
with $\mathrm{d.o.f.}$ stands for degree of freedom. The uncertainty of the estimated parameters $\hat{\boldsymbol{\beta}}$ is then quantified based on uncertainty propagation law:
\begin{gather}
    \hat{\mathbf{Q}}_{\hat{\boldsymbol{\beta}}} = \hat{\sigma}_{0}^{2}(\mathbf{A}^{\text{T}}(\mathbf{B}^\text{T}\mathbf{P}^{-1}\mathbf{B})^{-1}\mathbf{A})^{-1}.
\end{gather}
Therefore, the estimated uncertainty of $\hat{\boldsymbol{\beta}}$ depends on the observation uncertainties $\mathbf{Q}_z$, the predefined functional relationship (Eq.~\ref{eq:lsq_function}) and also its Taylor expansion (Eq.~\ref{eq:Talyor}). The first term is the aleatoric part of the total uncertainty and rigorously considered in a least-square estimation. The last two factors together form the epistemic part of the total uncertainty, although their impacts are usually argued to be small by assuming the pre-defined relationship Eq.~\eqref{eq:lsq_function} is accurate and all the "$+0$" term to be zero-mean Gaussian. It is clear that the aleatoric and epistemic uncertainties are coupled in the uncertainty propagation approach.

\subsubsection{Deep learning view}
\label{sec:Deep learning view}
To show the uncertainty sources in a deep learning regression model, we start the discussion with a relatively simple case. Following the definitions in the previous sections, $\mathbf{X}=[\mathbf{x}_1,\dotsc,\mathbf{x}_n]^\mathrm{T}$ denote the input dataset, and $\mathbf{Y}=[\mathbf{y}_1,\dotsc,\mathbf{y}_n]^\mathrm{T}$ denote the corresponding target dataset. The pair $\mathcal{D}=\{\mathbf{X}, \mathbf{Y}\}$ denotes the training dataset of $n$ samples. Considering a linear relationship constrained by the regularization parameter $\lambda$ (\citealt{khalaf2005Ridge}, so-called Ridge regression or L2-regularization), we can find the analytical solution:
\begin{equation}
    \boldsymbol{\hat{\theta}} = \left(\mathbf{X}^\mathrm{T}\mathbf{X} + \lambda\boldsymbol{\mathbf{I}}\right)^{-1}\mathbf{X}^\mathrm{T}\mathbf{Y},
\end{equation}
which is equivalent to finding the mode of the full posterior distributions~\citep{deisenroth2020Math4ML}:
\begin{equation}
    \arg\min_{\boldsymbol{\theta}}\sum_{i=1}^n\left(y_i - \boldsymbol{\theta}^{\mathrm{T}}\mathbf{x}_i\right)^2+\lambda\lVert\boldsymbol{\theta}\lVert_2^2 \ \equiv \arg\max_{\boldsymbol{\theta}}P\left(\boldsymbol{\theta}\right)\prod_iP\left(y_i|\mathbf{x}_i,\boldsymbol{\theta}\right).
\end{equation}

Assuming the prior distributions of parameters are independent Gaussian distributions $p(\boldsymbol{\theta})=\mathcal{N}(0, \sigma_p\mathbf{I})$, and the likelihood are conditionally independent $p(y_i|\boldsymbol{\theta},x_i)=\prod_iP\left(y_i|\mathbf{x}_i,\boldsymbol{\theta}\right)=\prod_i\mathcal{N}(y_i;\boldsymbol{\theta}^\mathrm{T}\mathbf{x}_i,\sigma_n^2)$, the inference given a new set of inputs $\mathbf{x}^*$ can be analytically expressed as:
\begin{equation}
    p(y^*|\mathbf{x}^*,\hat{\boldsymbol{\theta}}) = \mathcal{N}(y^*;\hat{\boldsymbol{\theta}}^T\mathbf{x}^*,\sigma_n^2),
\end{equation}
indicating that $y^*$ is sampled from a Gaussian distribution with the expectation of $\hat{\boldsymbol{\theta}}^T\mathbf{x}^*$ and variance of $\sigma_n^2$. The most plausible mode estimation of $f^*=\hat{\boldsymbol{\theta}}^\mathrm{T} \mathbf{x}^*$ is associated with inherent random noise $\sigma_n$ (aleatoric uncertainty). Since the prediction of a Ridge regression focuses on the mode rather than providing the full posterior distribution, it provides no information on the epistemic uncertainty.

To consider the distribution of trainable parameters $\boldsymbol{\theta}$ based on Bayesian statistics, we estimate the whole posterior distributions $p\left( \boldsymbol{\theta}|\mathcal{D}\right)$ instead of the mode $\hat{\boldsymbol{\theta}}$. The posterior distribution of the trainable parameters $\boldsymbol{\theta}$ can be expressed using Bayes' theorem:
\begin{equation}
    p\left(\boldsymbol{\theta}|\mathcal{D}\right) = \frac{p\left(\mathcal{D}|\boldsymbol{\theta}\right)p(\boldsymbol{\theta})}{p(\mathcal{D})}.
\end{equation}
In this way, we generalize the ridge regression to Bayesian linear regression (BLR) and also consider the discrepancies among different parameters given the same set of training samples (epistemic uncertainties). It is worth noting that a BLR model is nothing else but a Gaussian process regression (the same as the well-known Kriging or Collocation in the geoscience community) with a linear kernel~\citep{deisenroth2020Math4ML}. By inference, the predictive distribution of $\mathbf{y}^*$ is given by:
\begin{equation}
    p\left(y^*|\mathbf{x}^*, \boldsymbol{\theta}\right) = 
    \int p\left(y^* | \mathbf{x}^*, \boldsymbol{\theta}\right) p\left( \boldsymbol{\theta}|\mathcal{D}\right) d\boldsymbol{\theta} = \mathcal{N}\left(\boldsymbol{\hat{\theta}}^\mathrm{T}\mathbf{x}^*, \mathbf{x}^{*\mathrm{T}}\Sigma\mathbf{x}^{*} + \sigma_n^2\right),
    \label{eq:Inference_MAP_Gaussian}
\end{equation}
from which we can obtain the uncertainties associated with the model and data. The first part, $p\left(\mathbf{y}^* | \mathbf{x}^*, \boldsymbol{\theta}\right)$, describes the prediction, given the input data and the model, whereas the second part, $p\left( \boldsymbol{\theta}|\mathcal{D}\right)$, describes the model distribution, given the data. The variances of different possible parameters, i.e., the epistemic uncertainty, is described by $\Sigma=\left(\mathbf{X}^\mathrm{T}\mathbf{X} + \sigma_n^{-2}\boldsymbol{\mathbf{I}}\right)^{-1}$. At this stage, it is evident that this epistemic uncertainty is coupled with input data and the associated uncertainty (aleatoric) and, therefore, challenging to be fully separated.

So far, we have attempted to clarify the sources of both aleatoric and epistemic uncertainties in a BLR model. However, we have made three strong, and practically uncommon assumptions: (i) The input-target relationship is linear; (ii) All the relevant distributions are Gaussian; (iii) The aleatoric uncertainties are homoscedastic, namely independent of inputs. We briefly introduce the solutions for reducing these three assumptions in the following parts of this section, followed by a detailed discussion and literature review in Section~\ref{sec:UQ-DL}.

To tackle the first issue, we move from BLR to Bayesian deep learning (BDL)\footnote{Some debate exists about whether BDL algorithms are fully Bayesian or just probabilistic. We use the term BDL without implying the former, to remain consistent with existing literature.}, in which the linear regression is replaced by a neural network~\citep{lecun2015DL,goodfellow2016DL}:
\begin{equation}
f(\mathbf{x};\boldsymbol{\theta})=\varphi_L\left(\varphi_{L-1}\left(\cdots\varphi_1\left(\mathbf{x}\right)\right)\right),
\end{equation}
where $\varphi_L$ denotes the projection of the $L^\mathrm{th}$ layer which is typically composed of trainable parameters $\boldsymbol{\theta}_L$ (weights and biases) and non-linear activation functions. It is worth noting that the prior distributions of $\boldsymbol{\theta}_L$ are typically assumed as isotropic Gaussian distributions without credible prior information~\citep{fortuin2021bayesian}, meaning that the derived posterior distribution may not strictly follow the full Bayesian statistics. Therefore, the whole approach may be better understood as a pragmatic, Bayesian-inspired computation. In principle, such a neural network can approximate any non-linear relationship accurately~\citep{hornik1989UniversalApproximator}, but the computational complexity will increase remarkably.

By generalizing from Gaussian distribution to all possible distributions (second issue), a crucial problem arises that the posterior distribution $p(\boldsymbol{\theta}|\mathcal{D})$ cannot be computed analytically, and therefore, the second equal sign in Eq~\eqref{eq:Inference_MAP_Gaussian} does not hold anymore. This issue brings the challenge of determining the epistemic uncertainty associated with such neural networks. To this end, we need to approximate the complicated distributions by simpler distributions, which is known as variational approximation~\citep{blei2017VIreview}. A huge amount of effort has been dedicated to solving this issue to obtain realistic epistemic uncertainties by considering different requirements~\citep{abdar2021UQinDLsurvey,gawlikowski2023UQinDLsurvey}, which we will discuss in detail focusing on the ones have been applied to ECV studies (Section~\ref{sec:UQ-DL}).

Finally, the aleatoric uncertainties usually depend on the corresponding input data uncertainties, as discussed in Section~\ref{sec:Statistical view}. To this end, the heteroscedastic aleatoric uncertainties (issue 3) are crucial for ECVs, given the varying quality of input data. In a typical BDL pipeline, the heteroscedastic aleatoric uncertainties can be estimated by adding an output of the dedicated neural network to represent the associated variance of the predictions~\citep{kendall2017WhatUncertainties}:
\begin{align}
    p(\mathbf{y}^*|\mathbf{x}^*, \mathcal{D}) &\sim \mathcal{N}(\mu\left(\mathbf{x}^*;\boldsymbol{\theta}_1\right), \sigma^2_\mathrm{ale}(\mathbf{x}^*; \boldsymbol{\theta}_2)),\label{eq:Distribution_NLL}\\
    \boldsymbol{\mu}\left(\mathbf{x}^*;\boldsymbol{\theta}_1\right) &= f\left(\mathbf{x}^*; \boldsymbol{\theta}_1\right), \\
    \boldsymbol{\sigma}^2_\mathrm{ale}(\mathbf{x}^*; \boldsymbol{\theta}_2) &= \exp(f\left(\mathbf{x}^*; \boldsymbol{\theta}_2\right)),
\end{align}
where $\boldsymbol{\mu}$ and $\boldsymbol{\sigma}^2_\mathrm{ale}$ are the two outputs of the same neural network $f$ but parameterized by different parameters $\boldsymbol{\theta}_1$ and $\boldsymbol{\theta}_2$. The differences may only be present in the output layer. The positive values of the estimated variances are ensured by $\exp(*)$, which can be replaced by other positive functions. By assuming a Gaussian likelihood, we formulate the negative log-likelihood loss function as:
\begin{align}
    \mathcal{L}(\boldsymbol{\theta}) &= -\log p_\theta(\mathbf{Y}|\mathbf{X}, \boldsymbol{\theta})\nonumber \\ 
    &= \frac{1}{2}\log(\boldsymbol{\sigma}^2_\mathrm{ale}) + \frac{1}{2}
    (\boldsymbol{Y}-\boldsymbol{\mu})^T{\rm diag}(\boldsymbol{\sigma}^{-2}_\mathrm{ale})
    (\boldsymbol{Y}-\boldsymbol{\mu})
    + \mathrm{constant},
    \label{eq:Loss_NLL}
\end{align}
but can also be derived based on other tractable distributions~\citep{nair2022NLL_Laplacian,Kiani2024Laplacian}. With this modification, the training input uncertainties are considered (although only implicitly), especially when the training dataset covers a reasonably large partition of the whole data distribution. However, certain limitations exist. First, the covariances cannot be estimated since the large number of parameters makes all but the sparsest covariance matrices intractable. Second, the implicit consideration of input uncertainties may result in unrealistic uncertainties of the outputs, and the associated uncertainty of $\mathbf{x}^*$ during inference is hardly considered since the same input values may be associated with totally different uncertainties. This issue is critical for ECV studies when dealing with real-world sEO data. Some alternative ways to quantify the aleatoric uncertainties based on the available uncertainty information of inputs will be discussed in Section~\ref{sec:DEMC}.
\section{Uncertainty quantification in deep learning}
\label{sec:UQ-DL}
\subsection{Methods for solving the posterior distribution of parameters}
\label{sec:UQ_DL_Intro}
As described in the previous section, the posterior model distribution $p(\boldsymbol{\theta}|\mathcal{D})$ cannot be expressed analytically without linear relationship and Gaussian assumptions. The preliminary task in Bayesian deep learning is to find a way to approximate the posterior distribution to the best possible extent. One of the most rigorous ways of this approximation task is to use a posterior distribution based on variational parameters with simpler distributions, such as Gaussian. This approach is known as variational inference (VI) and is widely used in the machine learning community~\citep{blei2017VIreview}. Due to computational complexity, various studies are dedicated to achieving similar performance without explicitly finding variational parameters~\citep{wilson2020BDL,jospin2022BNNTutorial}. The categorizations of different BDL algorithms might vary slightly within existing literature~\citep{abdar2021UQinDLsurvey,gawlikowski2023UQinDLsurvey}. We choose to categorize the different BDL methods into the following four classes and describe each in detail in the following subsections:
\begin{itemize}
    \item \textbf{Bayesian neural network (BNN)} refers to neural networks with all parameters associated with distributions instead of only modes. These distributions can be explicitly approximated by gradient-based optimization or fitting locally based on tractable distributions like Gaussian. The detailed techniques and applications to ECVs are listed in Section~\ref{sec:Bayesian neural networks}.
    \item \textbf{Sampling methods} refer to the approaches that directly sample the output distributions but do not approximate the distributions of model parameters. Although the model parameters are still associated with certain distributions, they are not explicitly expressed. Section~\ref{sec:Sampling methods} introduces Monte Carlo sampling and also the special adaptions based on the characteristics of neural networks. These methods do not require retraining multiple individual deep learning models.
    \item \textbf{Ensemble methods} refer to the approaches that require training and aggregating multiple individual deterministic models. The distributions of model parameters are not explicitly estimated. However, the variance of the outputs can be quantified by considering the differences among different ensemble members. Some of the widely-used ensemble methods are introduced in Section~\ref{sec:Ensemble approaches}.
    \item \textbf{Other methods} include additional methods with special modifications for specific goals, especially for modern developments towards large models with big data, see Section~\ref{sec:Other advancements in the era of large models}. They may have some overlap with the first three categories.
\end{itemize}

\subsection{Bayesian neural networks}
\label{sec:Bayesian neural networks}
Contrary to classical deep learning neural networks, which estimate the modes of each trainable parameter during training, a BNN attempts to model the full posterior distribution $p\left(\boldsymbol{\theta}|\mathcal{D}\right)$ of all trainable parameters $\boldsymbol{\theta}$ given a set of training data $\mathcal{D}$. As mentioned in Section~\ref{sec:UQ_DL_Intro}, the posterior distribution $p(\boldsymbol{\theta}|\mathcal{D})$ cannot be computed analytically, especially for highly non-linear deep learning models and when the parameters have high dimensions. The idea is to approximate the posterior distribution by some simpler and tractable distributions $q_{\boldsymbol{\lambda}}(\boldsymbol{\theta})$ (called variational distributions), such as the Gaussian distribution, characterized by another set of parameters $\boldsymbol{\lambda}$. To achieve this goal, we try to maximize the similarity between the two distributions. In practice, we may minimize the Kullback-Leibler (KL) divergence~\citep{kullback1951KLDivergence} by adjusting $\boldsymbol{\lambda}$:
\begin{equation}
\mathrm{KL}\left(q_{\boldsymbol{\lambda}}(\boldsymbol{\theta})||p(\boldsymbol{\theta}|\mathcal{D})\right)=\int q_{\boldsymbol{\lambda}}(\boldsymbol{\theta})\log\frac{q_{\boldsymbol{\lambda}}(\boldsymbol{\theta})}{p(\boldsymbol{\theta}|\mathcal{D})}d\boldsymbol{\theta}.
    \label{eq:KL_divergence}
\end{equation}
The KL divergence is a measure of how similar two probability distributions are, and is frequently used as a loss function for VI in machine learning. 
In the following subsections, we discuss how to adapt the KL divergence and incorporate it into the optimizing process of a neural network.

\subsubsection{Bayes by Backprop}
\label{sec:Bayes_by_Backprop}
Bayes by Backprop~\citep{blundell2015BayesBackprop} is a famous adaption of VI to deep neural networks. To find the variational distribution that can minimize the KL divergence (Eq.~\eqref{eq:KL_divergence}) under the constraint of given dataset $\mathcal{D}$, we can rearrange the optimization problem into evidence lower bound (ELBO) maximization:
\begin{equation}
    \arg\min_{q\in Q}\mathrm{KL}\left(q_{\boldsymbol{\lambda}}||p(\boldsymbol{\theta}|\mathcal{D})\right)=\arg\max_{q\in Q}\underbrace{\left\{\mathbb{E}_{\boldsymbol{\theta}\sim q}\left[p(Y|\boldsymbol{\theta})\right] - \mathrm{KL}\left(q_{\boldsymbol{\lambda}}||p(\boldsymbol{\theta})\right)\right\}}_{\mathrm{ELBO,} \mathcal{L}(q)}\;,
\end{equation}
which saves us from solving the posterior distributions. If we consider the distribution family $Q$ as Gaussian, we can formulate the targeted variational distributions as $q(\boldsymbol{\theta}|\boldsymbol{\lambda})=\mathcal{N}\left(\boldsymbol{\theta};\boldsymbol{\mu}, \boldsymbol{\Sigma}\right)$, where $\boldsymbol{\lambda}$ indicates the trainable parameters that should be optimized during backpropagation. Instead of estimating $\boldsymbol{\theta}$ during training, we estimate $\boldsymbol{\mu}$ and $\boldsymbol{\Sigma}$ on $\mathcal{D}$. The mean vector and covariance matrix may be estimated using the re-parameterization trick~\citep{kingma2015VariationalDropout} that detaches the $\boldsymbol{\mu}$ and $\boldsymbol{\Sigma}$ from the randomness in the approximate posterior. Hence, we model the parameters via
\begin{equation}
    \boldsymbol{\theta} = \boldsymbol{\mu} + \boldsymbol{\sigma}\odot\boldsymbol{\epsilon},
\end{equation}
where $\boldsymbol{\sigma}$ denotes the diagonal of $\boldsymbol{\Sigma}^{1/2}$ and $\boldsymbol{\epsilon}\sim\mathcal{N}(\mathbf{0}, \mathbf{I})$. Then, we can reformulate ELBO into a target loss function with respect to $\boldsymbol{\lambda}$ with the gradients as:
\begin{equation}
    \nabla_{\boldsymbol{\lambda}} \mathcal{L}(\boldsymbol{\lambda}) = \nabla_{\boldsymbol{\sigma},\boldsymbol{\mu}} \mathbb{E}_{\boldsymbol{\epsilon}\sim\mathcal{N}(0, \mathbf{I})}\left[\log p(Y|\boldsymbol{\mu} + \boldsymbol{\sigma}\odot\boldsymbol{\epsilon})\right] - \nabla_{\boldsymbol{\sigma},\boldsymbol{\mu}}\mathrm{KL}\left(q_{\boldsymbol{\sigma},\boldsymbol{\mu}}||p(\boldsymbol{\theta})\right),
\end{equation}
which has an analytical solution. Therefore, we can find the optimal solution using classical optimization algorithms, such as Adam~\citep{kingma2014Adam}. After training a BNN, we approximate the final posterior distribution $p\left(\mathbf{y}^*|\mathbf{x}^*, \mathcal{D}\right)$ by using sampling methods, namely drawing $M$ sets of weights  from the weights distribution and make the final predictions:
\begin{align}
   p\left(y^*|\mathbf{x}^*, \mathcal{D}\right)&=\mathbb{E}_{\boldsymbol{\theta}\sim q(\boldsymbol{\theta}|\boldsymbol{\lambda})}\left[p(\mathbf{y}^*|\mathbf{x}^*,\boldsymbol{\theta})\right]\\
   &\approx\frac{1}{M}\sum_{m=1}^Mp\left(y^*|\mathbf{x}^*,\boldsymbol{\theta}^{(m)}\right).
   \label{eq:integral_approx}
\end{align}

\subsubsection{Laplace approximation}
In Laplace approximation~\citep{mackay1992LaplaceApproximation}, we also approximate the posterior distribution $p(\boldsymbol{\theta}|\mathcal{D})$ using a Gaussian distribution around the true posterior using a second-order Taylor expansion around the posterior mode $\widehat{\boldsymbol{\theta}}$: 
\begin{equation}
    \log p(\boldsymbol{\theta}|\mathcal{D}) \approx
    \log p(\widehat{\boldsymbol{\theta}}|\mathcal{D}) + 
    \frac{1}{2}\left(\boldsymbol{\theta}-\widehat{\boldsymbol{\theta}}\right)^T\left(\mathbf{H}+\epsilon\mathbf{I}\right)
    \left(\boldsymbol{\theta}-\widehat{\boldsymbol{\theta}}\right),
\end{equation}
where $\mathbf{H}$ is the Hessian matrix of $\log p(\boldsymbol{\theta}|\mathcal{D})$. The posterior distribution can be expressed as:
\begin{equation}
    p(\boldsymbol{\theta}|\mathcal{D}) = \mathcal{N}\left( \widehat{\boldsymbol{\theta}},\left(\mathbf{H}+\epsilon\mathbf{I}\right)^{-1}\right).
\end{equation}

Inverting the Hessian matrix is computationally very heavy, and therefore the Laplacian approximation is often applied to the last layer of the network \citep{kristiadi2020beingBayeisan}.

\subsubsection{Applying BNN to ECVs}
The value of BNN has been widely realized due to its relatively solid theoretical background. Prior to the developments in approximating the posterior distributions using gradient-based optimization algorithms, the approximation of posterior distributions primarily relied upon methods such as Laplacian approximation. \cite{khan2006BNN4Runoff} designed a BNN based on multilayer perception (MLP) to model the rainfall-runoff relationship with uncertainty quantification. Moreover, they highlighted that the BNN outperformed the conventional MLP model with less severe overfitting issues because of considering parameter uncertainties. Following studies applied similar strategies to model other ECVs including groundwater level~\citep{maiti2014LA4GW}.

In recent years, Bayes by Backprop and its variants~\citep[see, e.g.,][]{liu2016SteinVariational} have been gaining more attention since they can be efficiently solved by existing gradient-based optimization techniques. Moreover, such methods are easier to be generalized to large data volumes and fit the current direction of developments towards "big data" and foundation models~\citep{li2023BigDataEarth}. Various studies employed this technique to estimate ECVs and achieved superior accuracies with realistic uncertainty information. \cite{clare2022BNN4OceanDynamics} applied BNN to classify dynamical ocean regimes and connect their uncertainty analysis with explainable artificial intelligence (xAI) techniques to improve the trustworthiness of proposed models. \cite{mo2022BCNN} and \cite{uz2022bridging} both employed Bayesian convolutional neural networks (BCNN) to bridge the one-year gap between GRACE and GRACE-FO missions to provide a continuous record of global TWS anomalies. The realistic predictive uncertainties are provided together with satisfactorily high-quality predictions. However, the trends reconstructed by BCNN are relatively less accurate than other methods such as deep convolutional auto encoders~\citep{uz2022bridging}, indicating the issue that BNN may have lower generalizability than other simpler proxies under domain shift~\citep{izmailov2021BNNPrior}. The short-term variability of the sea surface temperature was satisfactorily predicted by using BNN~\citep{luo2022BNN4SeaTemperature}. With the quantified uncertainties inherent in the data and caused by model itself, reasonable risk assessment can be performed while using the temperature prediction for forecasting extreme events. \cite{kang2024BNN4Discharge} applied BNN to estimate the relationship between source zone metrics and mass discharge, outperforming the existing upscaled models with fewer parameters and providing uncertainty information. Such an uncertainty-aware model holds great potential for risk assessments and decision making. A similar model was employed by \cite{xu2021BDL4Precipitation} for forecasting precipitation. They highlighted that the predictand data uncertainty might be ignored in a typical BNN pipeline and, therefore, proposed to jointly model the input/target data uncertainties and model uncertainties. This method is crucial towards a more realistic uncertainty quantification as we described in Section~\ref{sec:Deep learning view} and will be further discussed in Section~\ref{sec:DEMC}. A common finding from all the mentioned studies is that the accuracy and computational costs of BNNs critically depend on the pre-defined prior distributions~\citep{fortuin2021bayesian,jospin2022BNNTutorial}, which need to be carefully designed in advance.

\subsection{Sampling methods}
\label{sec:Sampling methods}
Although approximating the full posterior distributions based on VI provides the most rigorous information on the distribution of parameters, certain limits exist, including the computational cost, optimization difficulty, and sensitivity to prior distributions. Alternatively, we can sample from the posterior distribution without explicitly estimating it using Monte Carlo simulation. By combining these principles with deep neural networks, unique opportunities are enabled that can be exploited for the purpose of uncertainty quantification.

\subsubsection{Markov Chain Monte Carlo}
Markov Chain Monte Carlo (MCMC, \citealt{neal1992MC4Network,neal2011mcmc}), is a powerful technique used in statistics and various other fields to sample from complex probability distributions. It allows us to "explore" these distributions by generating sets of random samples that "resemble" the true underlying distribution, regardless of the complexity of this distribution. The sampling techniques are practical for approximating the predictive distribution of possible neural network parameters since the integral of the predictive distribution is often intractable. 

MCMC relies on Markov chains~\citep{bishop2006pattern}, which are sequences of random values where each value depends only on the previous one. MCMC constructs a chain that "walks through the landscape" set out by the desired distribution. Each step involves proposing a new value based on the current one and accepting or rejecting it probabilistically. The "walk" continues until the chain reaches equilibrium, meaning the distribution of values visited resembles the target distribution. Keeping track of the accepted values after reaching equilibrium allows us to get samples that reflect the target distribution, even if we could not directly sample from it:
\begin{equation}
    \boldsymbol{\theta}_t \sim q\left(\boldsymbol{\theta}_t|\boldsymbol{\theta}_{t-1}, \mathcal{D}\right).
\end{equation}

The number of transitions needs to be performed $T$ times, which must be determined in advance. Once we determine a sequence of weights $\boldsymbol{\theta}_1\dots\boldsymbol{\theta}_T$, the prediction reads:
\begin{equation}
    p\left(y^*|\mathbf{x}^*, \mathcal{D}\right) \approx \frac{1}{T}\sum_{t=1}^T p\left(y^*|\mathbf{x}^*, \boldsymbol{\theta}_t\right).
\end{equation}

However, the optimal value of $T$ is usually unknown and needs to be determined based on experiments. Moreover, the computational cost for MCMC is very high, especially for today's modern deep neural networks. Even though MCMC is the gold standard for sampling from complex distributions, they are not tractable for deep learning networks with billions of parameters. The huge size of trainable parameters prohibits saving the full set of parameters in a Markov chain, whereas modern big datasets cannot be processed at once but require stochastic training strategies.

\subsubsection{Monte Carlo dropout}
\label{sec:MC-Dropout}
Monte Carlo dropout (MC-Dropout; \citealt{gal2016MCdropout}) is a special application of Monte Carlo methods to deep neural networks based on the dropout mechanism~\citep{srivastava2014dropout}. The dropout mechanism can be expressed as:
\begin{equation}
    \boldsymbol{\theta}^{(i)} = \mathbf{s}^{(i)} \odot \boldsymbol{\theta}, \quad \mathbf{s}^{(i)} \sim \mathrm{Bernoulli}(\mathbf{s}|\alpha),
    \label{eq:MCDropout}
\end{equation}
where $\alpha$ denotes the probability that a node is activated and $\boldsymbol{\theta}$ denotes all trainable parameters in the model architecture. During each iteration, only this subset of the model parameters will be activated during the forward pass and optimized during backpropagation. Dropout is an efficient technique to mitigate the overfitting issue.

The principle of MC-Dropou is simple: we leverage the dropout also during inference, i.e., randomly selected nodes are deactivated to generate different outputs. The approach will be repeated $T$ times, and the final output of the network is computed as Eq.~\eqref{eq:integral_approx}. We can understand this approach as sampling from the full posterior distribution of $\boldsymbol{\theta}$ based on Eq.~\eqref{eq:MCDropout}. Although we cannot explicitly express the distribution of $\boldsymbol{\theta}$, we can directly sample the final outputs from the marginal distribution, namely:
\begin{equation}
    f(y^*|\mathbf{x}^*, \mathcal{D}) = \frac{1}{T}\sum_{t=1}^T\mathrm{NN}_{\boldsymbol{\theta}^{(i)}}(\mathbf{x}^*),
\end{equation}
where $\mathrm{NN}_{\boldsymbol{\theta}^{(i)}}$ denotes the neural network with a subset parameters $\boldsymbol{\theta}^{(i)}$ activated.

A drawback of this approach is that the model architectures vary during inference (see also Section~\ref{sec:Ensemble of different model architectures}), so the possible parameter distributions are coupled with the possible model distributions, although both of them are understood as epistemic uncertainties. Additionally, employing dropout requires more careful hyperparameter tuning to avoid potential degradation of performance~\citep{garbin2020DropoutandBN}.

\subsubsection{Stochastic weight averaging}
Stochastic weight averaging (SWA; \citealt{izmailov2018SWA}) exploits the randomness in the optimization algorithms (stochastic gradient descent; SGD).
During training, we often reduce the learning rate after certain iterations, and the model solution does not move much in the loss landscape. SWA with a Gaussian (SWAG; \citealt{maddox2019SWAG}) treats the model samples from the SGD as a Gaussian. Hence, SWAG fits a Gaussian distribution to the local geometry of the posterior:
\begin{equation}
    p(\boldsymbol{\theta}|\mathcal{D}) =\mathcal{N}\left(\boldsymbol{\mu}(\mathcal{D}), \boldsymbol{\Sigma}(\mathcal{D})\right),
\end{equation}
where
\begin{align}
    \boldsymbol{\mu}(\mathcal{D}) &= \frac{1}{N}\sum_{n=1}^N\boldsymbol{\theta}_n,\\
    \boldsymbol{\Sigma}(\mathcal{D}) &= \frac{1}{N}\sum_{n=1}^N\left(\boldsymbol{\theta}_n - \boldsymbol{\mu}(\mathcal{D})\right)\left(\boldsymbol{\theta}_n - \boldsymbol{\mu}(\mathcal{D})\right)^T
\end{align}
and $\boldsymbol{\theta}_1,\dotsc, \boldsymbol{\theta}_N$ is the model parameters from the $N$ last iterations. The assumption behind this approach is that the optimizer reaches a reasonably well-established local optimum and only oscillates around it. As a result, the trained parameters $\boldsymbol{\theta}$ represent the targeted relationship equally well.

\subsubsection{Applying sampling methods to ECVs}
Sampling methods have long been applied to approximate the posterior distributions due to their simplicity of implementation. MCMC and its derivatives such as evolutionary MC~\citep{liang2001EMC} are widely used for marginalizing BDL methods already in earlier epochs. The MCMC algorithms are usually combined with classical MLP models and compared to the deterministic version of the same model. \cite{kingston2005MCMC4Salinity} trained a MCMC MLP to forecast river salinity in the Murray River at Murray Bridge, South Australia. Similarly, \cite{zhang2009EMC4Streamflow} applied MCMC to quantify the uncertainties of steamflow simulations with a specific focus on discussing different treatments of uncertainties, and highlighted the importance of considering variable model architecture and informative prior knowledge for realistic uncertainty quantification. Both of the two mentioned studies reported that the proposed BDL models performed worse than deterministic MLP on the training set but showed superiority in the validation or test set, indicating the difficulty of optimizing a BDL model but also its robustness against overfitting issues. \cite{jana2008MCMC4Soil} approximated fine-scale soil hydraulic properties in two regional basins based on coarse-scale inputs and also reported better generalizability of MCMC MLP. The uncertainties quantified through MCMC simulations are mostly epistemic, whereas the uncertainties associated with sources other than the model parameters (mostly aleatoric) need to be additionally considered \citep{humphrey2016MCMC4Streamflow}. Building upon the concept of predictability, \cite{kiani2024bahamas} designed an autoencoder-based architecture and used it alongside Hamiltonian MC (HMC) to prove theoretically and empirically that the uncertainties derived are asymptotically convergent to the aleatoric uncertainties with convergence rate $N^{-p},~\frac{1}{2}<p\le{1}$, where $N$ is the number of MC samples used and $p$ proportional to the difference between two sequential components of the Markov chain. However, they noted that, in general, the convergence cannot be guaranteed, owing to the potential existence of outliers in the data. In recent years, the rapid growth in hardware capacities has enabled applying MCMC to more complex models such as LSTM for streamflow prediction~\citep{li2022MCMC4Steamflow}, targeting higher spatial resolution of ozone modeling~\citep{sun2021spatial}, or towards a global reconstruction of terrestrial water stroage~\citep{rateb2022MCMC4GRACE}. However, the potential for further scaling up the mentioned studies is still limited due to the fact that MCMC increases computational complexity~\citep{sun2021spatial}.

The standing-out disadvantage of MCMC concerning high computational cost has been noted in most of the above-mentioned studies. During the era of large models and big data, MC-Dropout has been increasingly used in practice. \cite{cobb2019MCDropout4Atmosphere} used MC-dropout to quantify the uncertainties associated with their atmospheric retrievals and highlighted the contribution of domain knowledge for achieving realistic uncertainties. Similarly, \cite{Tian_2022_subsurfacesalinity} applied MC-Dropout to quantify the uncertainties associated with their subsurface salinity estimations. However, they emphasized the pitfalls of the assessed uncertainties since the propagation mechanisms remain unclear. MC-Dropout can also be applied to more advanced model architectures without resulting in extremely high computational costs. For example, the coastal ocean time series, including oxygen, nutrients, and temperature, were predicted with satisfactory accuracy with realistic uncertainty information using LSTM by~\citet{Contractor_2021_nutrients}. MC-Dropout is also employed for a more complex model combining convolution and multi-head attention~\citep{gerges2024MCDropout4Wind} or convolutional autoencoder with skipping connections~\citep{uz2024MCDropout} without additional barriers.

\subsection{Ensemble approaches}
\label{sec:Ensemble approaches}
Ensemble approaches are based on a combination of several different models (called ensemble members). Although there are various ways in which models can be defined and combined~\citep{ganaie2022EnsembleReview}, three are widely used, namely stacking, bagging, and boosting. In stacking, several models are trained on the same data, and a final model combines the predictions of individual ensemble members. This category can be further separated into two cases, either the model architectures are the same, referring to the well-known deep ensembles~\citep{lakshminarayanan2017DeepEnsemble}, or combining different model architectures. In bagging, several deterministic models are trained on subsets of the same training dataset, and the final prediction is the average of individual predictions. In boosting, several models are added in a sequential manner, and each ensemble member attempts to correct the predictions of the previous members. In the following, we provide more information on each of these approaches.

\subsubsection{Ensemble of different model parameters}
\label{sec:Deep ensembles}
Ensemble of different model parameters can be categorized as stacking, where the architectures of the ensemble members are identical. In case a simple averaging based on a prescribed distribution for the input data is used as the final model (being a deep learning model), this type of stacking is referred to as deep ensembles \citep{lakshminarayanan2017DeepEnsemble}. A deep ensemble framework involves multiple neural networks (individual members) sharing the same architecture and optimized using the same strategy. The trainable parameters of each member are initialized stochastically and, therefore, the optimized parameters are different from the rest of the members, forming different basins of attractions, or local optima~\citep{wilson2020BDL}. The diversity among individual members is enough to quantify the epistemic uncertainties realistically~\citep{fort2019DeepEnsemble}. The heteroscedastic aleatoric uncertainties can be quantified following Eqs.~\eqref{eq:Distribution_NLL} to~\eqref{eq:Loss_NLL} for each ensemble member as:
\begin{align}
    \mathcal{L}(\boldsymbol{\theta}_i) &= -\log p_\theta(\mathbf{Y}|\mathbf{X}, \boldsymbol{\theta}_i)\nonumber \\ 
    &= \frac{1}{2}\log(\boldsymbol{\sigma}_i^2) + \frac{1}{2}
    (\boldsymbol{Y}-\boldsymbol{\mu}_i)^T{\rm diag}(\boldsymbol{\sigma}^{-2}_i)
    (\boldsymbol{Y}-\boldsymbol{\mu}_i) 
    + \mathrm{constant},
    \label{eq:Loss_deep_ensemble}
\end{align}
where $\boldsymbol{\mu}_i$ and $\boldsymbol{\sigma}_i^2$ are the mean and variance of the $i$-th ensemble member. After obtaining $M$ individual models, one can obtain the final prediction and uncertainty as:
\begin{align}
    \boldsymbol{\mu}_* &= \frac{1}{M}\sum_{m=1}^M\boldsymbol{\mu}_{\theta_m}\label{eq:DE_mu},\\
    \boldsymbol{\sigma}_* &= \sqrt{\frac{1}{M}\sum_{m=1}^M\left(\boldsymbol{\mu}_{\theta_m}^2+\boldsymbol{\sigma}_{\theta_m}^2\right)-\boldsymbol{\mu}_*^2}\label{eq:DE_sig}.
\end{align}

In practice, it is recommended to train the network to predict the logarithmic variance, $\mathbf{s}_i = \log(\boldsymbol{\sigma}^2_i)$, since this modifies the loss function to:
\begin{equation}
    \mathcal{L} = \frac{1}{2}\mathbf{s}_i + \frac{1}{2}\exp(\mathbf{-s}_i)(\boldsymbol{Y}-\boldsymbol{\mu}_i)^2, 
    \label{eq:Loss_deep_ensemble_logvar}
\end{equation}
which avoids a potential division by zero and is therefore more numerically stable \citep{kendall2017WhatUncertainties}.

Although deep ensembles have proven as an effective approximation of Bayesian marginalization, they might suffer from the lack of diversity among ensemble members. In this case, repulsive deep ensembles are used \citep{Dangelo_2021}, in which a 'repulsive' term such as $-|\boldsymbol{\theta}^{(i)}-\boldsymbol{\theta}^{(j)}|,~j=1,...,T$ is added to Eq. \eqref{eq:Loss_deep_ensemble}, ensuring that $\boldsymbol{\theta}^{(i)}$ and $\boldsymbol{\theta}^{(j)}$ ($i\ne j$) are as different as possible.

\subsubsection{Ensemble of different model architectures}
\label{sec:Ensemble of different model architectures}
The ensemble approach of stacking provides a helpful way of combining the predictions of various models. Since disparate architectures are used in this approach, the algorithm spans the model space to the extent supplied by the practitioner, which might enhance the performance of the deep learning algorithm. The general approach is first to choose the ensemble models and then train them on the same dataset. Finally, the predictions of individual ensemble models are combined to provide the final prediction and its uncertainty, either through a predefined rule (e.g., averaging or maximum voting) or algorithms dedicated to finding the optimal architecture (such as neural architecture search, \citealt{herron2020NAS}). Considering $M$ models and denoting them with $\phi_{i,\boldsymbol{\theta}_i},~i=1,...,M$, each with parameters $\boldsymbol{\theta}_i$, and denoting the aforementioned combination and uncertainty computation rule or algorithm as $\Phi$, this ensembling approach can be mathematically represented as:
\begin{align}\label{General_formula_ensemble_different}
    \boldsymbol{\mu}_*, \boldsymbol{\sigma}_* = \Phi\left(\phi_{1,\boldsymbol{\theta}_1},\cdots,\phi_{M,\boldsymbol{\theta}_{m}}\right),
\end{align}

Compared to the previously mentioned deep ensembles in Section~\ref{sec:Deep ensembles}, an ensemble of different model architectures can explore larger distributions of all possible model parameters and, therefore, may result in better generalizability. However, the risk of mixing the model uncertainties and disparate model capacities exists and potentially results in higher epistemic uncertainties. Moreover, the hyperparameters of individual ensemble members need to be tuned individually, introducing additional workload.

\subsubsection{Ensemble of different datasets}
\label{sec:Ensemble of different datasets}
Creating ensembles of different datasets is categorized as bagging, although it encompasses boosting as well. This ensemble approach differs from stacking in that it uses the same model for all the ensemble members but trains each identical model on a bootstrapped version of the same dataset. The final prediction and its uncertainty are, respectively, the combination (voting or averaging) of the predictions of individual ensemble members and their spread. The mathematical representation of this ensembling approach is similar to the one presented in Eq. \eqref{General_formula_ensemble_different}, but the model architectures are identical, and the combination rule is simply voting or averaging. Examples of widely used models based on bagging include the so-called Random Forest and Extra Trees. In case ensemble members are added sequentially to correct the predictions of the previous ensemble member, the resulting ensembling approach is termed boosting. In training such models, more attention is paid to the predictions with the largest error to find the required number of ensembles that can predict the training data with arbitrary accuracy.

\subsubsection{Applying ensemble approaches to ECVs}
Deep ensembles are one of the most widely-used methods for quantifying uncertainties associated with ECVs derived using deep learning algorithms because of their straightforward implementation, superior performance, and even better generalization compared to BNN under domain shift~\citep{izmailov2021BNNPrior}. Using a collection of remote sensing data, \cite{Tollenaar_2024_icesheet} used an ensemble of CNN models to detect---with uncertainty---the areas where bare ice is exposed in Antarctica, which provide critical information on the evolution of the ice-sheet. \cite{Folino_2023_precipitation} applied deep ensembles to more robustly detect extreme precipitation events, such as those in southern Italy. Similarly, \cite{Sha_2024_weather} applied deep ensembles for the prediction of severe weather events in the United States. \cite{Andersson_2021_seaice} applied an ensemble of 25 members trained on climate simulations for 6-month-ahead sea ice concentration predictions, outperforming the state-of-the-art dynamical models. The quantified uncertainties are reliable and invaluable for stakeholders to adapt their activities accordingly. The principle of deep ensembles can also be applied to dynamically average different physical models, which is promising to overcome the deficiencies of individual models~\citep{sengupta2020DE_GeophysicalModels}. Some studies did not model the aleatoric uncertainties as described in Section~\ref{sec:Deep ensembles}, but rather simply retrained the same model multiple times and computed the average for a more robust estimation~\citep{lopez2021DE4cloud,Haynes_2024_upperatmospherictemperature}. Moreover, some studies tried to better modeling the aleatoric uncertainties based on input uncertainty information rather than allow the networks to generate by themselves. \cite{gou2024global} used deep ensembles in the context of data assimilation using deep learning models to enhance the spatial resolution of TWS anomalies derived from GRACE(-FO). This is achieved by incorporating hydrological data through a downscaling approach, resulting in a model that provides accurate and high-resolution TWS anomalies on a global scale. They modified the deep ensemble approach and combined it with MC simulations to account for aleatoric uncertainties caused by input data uncertainties, also during inference. However, they noted that not all the relevant input uncertainties might be captured and, therefore, the estimated uncertainties might be overly optimistic. This approach is further discussed in Section~\ref{sec:DEMC}. Furthermore, some studies attempted to benefit from multiple ensembling principles. \cite{liu2023DE4WindSpeed} trained multiple spatiotemporal models for wind speed forecasting and also trained based on different training datasets. They also let the model directly output associated uncertainties. Therefore, their approach is a combination of all the three types of ensembles mentioned in this section and achieves high-precision point prediction with reliable probabilistic information.

Ensemble of different model structures also shows benefits in various fields in geosciences~\citep{natras2022ensemble}, including in ECV studies. For instance, \cite{Jose_2022_ensemble_precipitation} used an ensemble of six different machine learning models to provide accurate daily forecasts of precipitation and temperature over the Western Ghats region in India. They demonstrated significant improvement compared to the forecasts based on traditional general circulation models. \cite{Du_2023_ensemble_landcover} applied five different ensemble models to map the land cover in the Xinjiang region in northwest China. They showed a better performance of their ensemble algorithm compared to that of the individual models, thereby providing incentive for the utilization of ensemble learning for land cover mapping in arid areas that are subject to increased attention under the ongoing climate change. Although the stacking approach has gained increasing attention, it might suffer from the same ailments as deep ensembles (i.e., lack of diversity among ensemble models), as well as being computationally expensive to implement. Moreover, the discrepancies among different ensemble members can be relatively large since the individual model architecture may have different capacities.

Bagging has also been applied to ECV studies but more often combined with machine learning models, such as tree-based models, rather than deep neural networks. \cite{heydari_2024_boosting_ensembling} analyzed the performance of several boosting models (including XGBoost and Gradient Boosting Machine) for water quality forecasts in Prespa Lake in Greece. They demonstrated the effectiveness of boosting in their study, which is in agreement with the general usefulness of boosting in hydrological problems \citep[see e.g.,][]{kermani_2021_boosting_ensembling}. \cite{malakouti_2023_boosting_ensembling} applied several bagging and boosting models for the prediction of global temperature change and showed that in this problem, Extra Trees (a bagging model) presented the best prediction performance. Hence, ensembling based on bagging or boosting is advantageous in ECV problems. Also, combinations of these approaches with stacking yield promising results. For example, \cite{zaier2010estimation} combined boosting and stacking based on multiple MLP models for more accurate lake ice thickness estimations.

\subsection{Explicitly considering data uncertainty in a deep learning model}
\label{sec:DEMC}
As described above and already mentioned in Section~\ref{sec:Introduction}, the majority of investigations about uncertainty quantification in deep learning models focus on epistemic parts but do not pay the same amount of attention to data quality~\citep{gruber2023SourcesOfUncertainty}. More specifically, the uncertainties caused by input data, especially during inference, are largely ignored. It is possible to dynamically weigh the loss values of individual samples or batches based on the available label uncertainty information~\citep{kiani2022inclusion} or based on certain criteria~\citep{gou2024DSOBP} for a more robust estimation, sharing the similar strategy as Eq.~\ref{eq:LSQ_weight_parameter} in the conventional statistical model. However, the predictive uncertainties caused by the input features are still missing. One of the reasons may be the fact that typical datasets for vision and language modeling tasks, which inspired the majority of deep learning developments, are not associated with uncertainty information~\citep[see, e.g., ][]{deng2009imagenet}. On the contrary, the classical statistical approaches (Section~\ref{sec:Statistical view}) commonly used by geoscience and climate scientists emphasize uncertainty propagation, namely the consequence of input uncertainties on the estimations based on a given model. As a result, applying mature and established uncertainty quantification approaches developed based on classical deep learning tasks, such as vision and language processing, to ECV studies may result in the loss of crucial information, especially considering the varying real-world situations and the inherent measurement noises of satellite systems~\citep{gawlikowski2023UQinDLsurvey}. It is relevant to explicitly consider the data uncertainty, whenever they are available, in a deep learning model for reaching realistic uncertainty information.

\cite{jungmann2024UPinNN} proposed an analytical solution to propagate aleatoric uncertainties caused by the inherent input uncertainties through a neural network by utilizing the law of uncertainty propagation and provided rigorous theoretical results. In theory, this method is computationally efficient since it only needs the parameters of the network, with the only prerequisite that the neural networks should be continuously differentiable. The non-fully-connected layers, such as convolution and pooling layers, need to be expressed as vector-matrix multiplication to ensure mathematical rigor. In practice, the differentials can be obtained by using automatic differentiation engines~\citep{tensorflow2015whitepaper,paszke2019pytorch}. As an alternative, we can benefit from sampling methods. Test-time augmentation~\citep{ayhan2018TestAugmentation} is a useful tool for quantifying the aleatoric uncertainties by applying augmentation principles to generate test samples. To this end, augmentation techniques are used to mimic the distribution of the test set so that the impacts of input uncertainties can be estimated. The motivation for this approach is that images are normally not associated with quantified uncertainties. Therefore, augmentation techniques are used to mimic the possible errors of the input images so that the output uncertainties can be quantified ~\citep{wang2019AleatoricUncertainty}. \cite{loquercio2020general} proposed a framework to explicitly consider the data uncertainties by propagating them using assumed density filtering~\citep{boyen2013ADF}. Then, the data uncertainties are combined with model uncertainties quantified using MC-Dropout to quantify the total uncertainty. The proposed method is not dependent on the network architecture and task, does not change the optimizing process, and can be applied to an already trained model.

Similar sampling strategies can be applied to overcome the limitation of the classical deep ensembles approach: the aleatoric uncertainties are directly generated by the networks without sensing the real uncertainty information of the input data $\mathbf{X}$. Therefore, the heteroscedastic uncertainty may differ from the real ones caused by the input errors. The limitation can be circumvented by combining deep ensemble and Monte Carlo sampling approach to better utilize the uncertainty information of inputs during inference~\citep{gou2024global}. The basic procedure remains the same as for classical deep ensembles (Section~\ref{sec:Deep ensembles}), but now the model gives one output, namely the predicted mean ($\boldsymbol{\mu}$). The individual ensemble members can be optimized based on the classical loss functions, which can simplify the loss landscape and may stabilize the optimization process. Once all the individual members are trained, Monte-Carlo simulations are performed during inference based on each of them. The first step is to sample from the distribution of the inputs based on their uncertainty information:
\begin{equation}
    \mathbf{x}_i^* \sim \mathcal{N}(\mathbf{x}^*,\boldsymbol{\sigma}_{\mathbf{x}^*}^2).
\end{equation}
Then, the Monte Carlo simulations are run $I$ times and quantify the aleatoric uncertainties as follows:
\begin{align}
    \boldsymbol{\mu}_{\boldsymbol{\theta}_m}^* &= \frac{1}{I}\sum_{i=1}^I\boldsymbol{\mu}^*_{\boldsymbol{\theta}_m,i} = \frac{1}{I}\sum_{i=1}^If(\mathbf{x}_i^*|\boldsymbol{\theta}_m),\label{eq:MCDE_mu}\\
    \boldsymbol{\sigma}_{\theta_m}^* &= \sqrt{\frac{1}{I}\sum_{i=1}^I(\boldsymbol{\mu}^*_{\boldsymbol{\theta}_m,i} - \boldsymbol{\mu}_{\boldsymbol{\theta}_m}^*)^2},\label{eq:MCDE_sig}
\end{align}
and obtain the final predictions and uncertainties by putting Eqs.~\eqref{eq:MCDE_mu} and~\eqref{eq:MCDE_sig} into Eqs.~\eqref{eq:DE_mu} and~\eqref{eq:DE_sig}. Quantifying the heteroscedastic uncertainties using MC simulations has a similar philosophy as a sensitivity analysis~\citep[see, e.g.,][]{hofer1999sensitivity,cacuci2004comparative,gou2023LOD} that are used in quantifying uncertainties of complex systems (Appendix~\ref{appendix:Sensitivity Analysis}). To this end, the epistemic uncertainty is quantified in the same way as with classical deep ensembles, namely exploring the different modes in the function space. However, the aleatoric uncertainty is now explicitly dependent on the input uncertainties, which are supposed to provide more realistic uncertainty information and better address the needs of handling sEO data.

\subsection{Other advancements in the era of large models}
\label{sec:Other advancements in the era of large models}
In previous sections, we described multiple widely used uncertainty quantification approaches. These approaches can be categorized into three classes, with the caveat that these are not necessarily the only methods available but are specifically highlighted due to the nature and purpose of the present survey. For a more comprehensive review of existing deep learning techniques for uncertainty quantification, we refer to topical survey papers, such as the ones by \cite{abdar2021UQinDLsurvey} or ~\cite{gawlikowski2023UQinDLsurvey}. In this section, we include some recent developments specifically designed for the era of large models with big data. They are becoming increasingly critical in the context of applying deep learning models to quantifying ECVs on a global scale and considering complex climate systems~\citep{schneider2023AI4ClimateModel,bi2023Pangu,zhang2023AI4Weather_Nature,lam2023Weather_Science_noUQ}. The recent efforts towards establishing Earth and climate foundation models also highly desire the realistic approaches for quantifying uncertainties~\citep{zhu2024FM,tuia2024AI4AEO,bodnar2024Aurora}. The potential for overcoming the difficulty of applying methods like MCMC to a larger scale due to future boost of computational resources has been expected in previous studies~\citep{kingston2005MCMC4Salinity} but has not come true since the modern developments are towards significantly larger models. The complexity of the model architectures and the data volume have been increasing dramatically, along with the evolution of computational resources. Therefore, approximating the full distributions of all trainable parameters (Section~\ref{sec:Bayesian neural networks}) or sampling the posterior distributions based on Markov Chain (Section~\ref{sec:Sampling methods}) are not affordable. Even training complete models several times to formulate deep ensembles becomes costly. With this background, multiple studies investigated the proxies of ensembles by only stochastically varying parts of the deep learning model to quantify epistemic uncertainties. To this end, de-correlated predictors that can sample the weight distributions may be obtained without training and storing copies of the complete networks and large training datasets in memory. \cite{turkoglu2022FiLM} introduced FiLM-Ensemble, which is based on the concept of feature-wise linear modulation to generate an ensemble to provide well-calibrated epistemic uncertainties implicitly. \cite{halbheer2024LoRA} proposed low-rank adaptations (LoRA) to replace the respective linear layers with custom LoRA modules and generate varying predictions. Both studies can be understood as an implicit ensemble, which can provide realistic epistemic uncertainties without the need to repeat training the whole model. Both studies have not been applied to ECV studies but have already been comprehensively examined on classical computer vision datasets, where they have shown potential for application to ECV-related studies. The diffusion probabilistic models~\citep{ho2020Diffusion} have recently gained numerous successes, and their probabilistic nature is beneficial for quantifying uncertainties in weather forecasting model~\citep{price2023Gencast} or land surface model~\citep{lu2024DBUQ}.
\section{Applications}
\label{sec:Applications}
In this section, we apply selected BDL approaches to two real-world ECV estimation use cases to discuss the resulting uncertainties. Since the previous survey on existing studies in Section~\ref{sec:UQ-DL} involves various ECV parameters and spreads across a wide range of the Earth system, quantitative discussion about different uncertainty quantification approaches is difficult. The two application cases in this section intend to provide straightforward feelings on the characteristics of the aforementioned methods. Section~\ref{sec:Application-Snow} discusses the results for SCF estimations based on deep ensembles and focuses on the fidelity of the estimated uncertainties compared to actual errors. Section~\ref{sec:Application-TWS} discusses the results for TWS change estimations based on four selected deep learning approaches introduced in Section~\ref{sec:UQ-DL} and focuses on the differences and similarities of resulting uncertainties. In addition, the uncertainties derived from least-squares estimation are also included for comparison.

\subsection{Snow cover}
\label{sec:Application-Snow}
Snow cover is recognized as an ECV due to its significant impact on the Earth's climate system and hydrological cycles. It affects the albedo, or reflectivity, of the Earth's surface, which in turn influences global and regional temperatures by reflecting or absorbing sunlight. Snow acts as a natural insulator and water reservoir, releasing freshwater during warmer months, crucial for drinking water, agriculture, and hydroelectric power. Changes in snow cover also affect weather patterns, ecosystem dynamics, and climate feedback mechanisms, making it integral to climate modeling and predictions. Information about the snow cover is important for several applications, including climate change~\citep{brown2011} and environmental monitoring~\citep{sturm2001}, water resource management~\citep{sturm2017}, and forecasting hazards such as avalanches~\citep{eckert2024} and floods~\citep{yan2023}. Various techniques for surveying the snow coverage on regional and global scales exist. Meteorological observations and regular manual surveys of snow depth and snow density are traditional methods to estimate the snow water equivalent and to track the evolution of the snow cover. Contrary to the meteorological observations, satellite images provide continuous spatial measurements, acquired globally at regular intervals. During the last decades a number of methods for snow-cover mapping have been developed for optical as well as for active and passive microwave sensors~\citep{hall2002, tsai2019, pulliainen2020}. Because the spatial resolution of the satellite images from today’s passive microwave sensors is coarse (5–25 km), and the active microwave sensors may struggle to provide reliable information about the snow~\citep{tsai2019}, optical images from sensors like MODIS, AVHRR and Sentinel-3 SLSTR are often applied in both regional and global snow monitoring \citep{hall2002, husler2014, solberg2021}.

The \href{https://climate.esa.int/en/projects/snow/}{ESA Snow$\_$cci project} aims to contribute to the understanding of snow in the climate system by generating consistent, high quality, long-term datasets that meet the requirements of the GCOS. The objective of Snow$\_$cci is to generate homogenized long-time series of daily global snow extent maps from optical satellite data and daily global snow water equivalent products from passive microwave satellite data. In addition, the project aims to generate corresponding, pixel-wise uncertainty products. To describe the snow extent, Snow$\_$cci uses the SCF variable. The methodology for estimating SCF is based on the SCAmod algorithm \citep{metsamaki2012}, which is based on a forward semi-empirical model where the at-satellite observed reflectance is expressed as a function of the fractional snow cover. For estimating the uncertainty, an error propagation approach has been applied for MODIS and SLSTR sensors \citep{salminen2018}.    

\subsubsection{Overview of data and models}
In the following, we present an application similar to the ESA Snow$\_$cci SCF estimation, but using a deep learning-based method to estimate SCF and the corresponding aleatoric, epistemic, and total predictive uncertainties. Our study area includes the Scandinavian Peninsula, where we define the area of interest to encompass Norway and Sweden, and the European Alps region. The data source for product generation is the Sentinel-3 Sea Land Surface Temperature Radiometer (SLSTR), which is based on the heritage of Envisat's Advanced Along Track Scanning Radiometer. SLSTR observes in two directions and has a swath width of \SI{1675}{\kilo\meter} (nadir) and \SI{750}{\kilo\meter} (backward). The sensor provides observations over Scandinavia at least once a day. The instrument has nine bands in the spectral range 0.55 to \SI{12}{\micro\meter}, but we consider only the bands S1 (\SI{554.27}{\nano\meter}), S2 (\SI{659.47}{\nano\meter}), S3 (\SI{868.00}{\nano\meter}), S5 (\SI{1613.40}{\nano\meter}) and S6 (\SI{2255.70}{\nano\meter}). We are not using bands S4 (\SI{1374.80}{\nano\meter}) and S7 (\SI{3742.00}{\nano\meter}) since we experienced, at several occasions, data of lower quality in these bands.

The Sentinel-2 Multi Spectral Instrument (MSI) is the main source of reference data as it is expected to be closest to ground truth due to its high-resolution data. Physics-based algorithms are used for retrieval, and the high resolution provides the ability to visually evaluate the results, including snow cover, by an experienced operator. The Sentinel-2 and Sentinel-3 images that were selected for training data generation were subject to automatic cloud detection and subsequent manual inspection to mark areas with erroneous cloud detection. We have chosen an accurate binary snow-cover algorithm~\citep{klein1998} to generate reference SCF maps from Sentinel-2 MSI. The binary result was checked against a pseudo-color image based on a combination of MSI bands making maximum contrast between snow-covered and snow-free areas by an image interpretation expert. The high-resolution image was then down-sampled by averaging to Sentinel-3 SLSTR resolution in order to be applied to train the U-Net for pixel-wise mapping of SCF from Sentinel-3 SLSTR data.

The data set was divided into training and test sets: The years 2016 and 2017 are used for training the deep learning model and the year 2018 is used for testing. To create the input-label pairs, the reference snow products were paired with Sentinel-3 images as follows:
\begin{itemize}
    \item Find SLSTR images that were recorded $\pm$24 hours compared to the Sentinel-2 reference image used to create the reference product,
    \item Remove pairs where the SLSTR image has an overlap of less than \SI{40}{\percent} of the reference product,
    \item Re-project the reference product onto the same grid as the SLSTR image using nearest neighbor interpolation and crop the SLSTR product to the rectangle enclosing the reference product with a 128 pixels margin,
    \item Apply automatic cloud detection to the SLSTR data~\citep{metsamaki2015} and mask out cloud pixels in the reference product by inserting a value representing ”unknown”,
    \item Remove any resulting pairs with less than 20 pixels not labeled as ”unknown”.
\end{itemize}

The model architecture used for single task problems is a slightly altered version of the U-Net proposed by~\cite{ronneberger2015}. The network consists of an encoder and a decoder, each of which consisting of a basic building block of three convolutional layers, with subsequent batch normalization and a ReLU-activation function. The convolutions are implemented with the padding-scheme proposed by \cite{liu2018}, called partial convolutions, to reduce edge effects. In the encoder part, the images go through four such blocks with down-sampling by a factor of two after each block. In the decoder part, the resulting feature maps are subject to four such blocks with a bi-linear up-sampling operation after each, also with a factor of two. The feature maps go through one more block with convolutions before the final two separate convolution layers that map to the desired snow cover fraction and aleatoric uncertainty (see Section~\ref{sec:Deep ensembles}), respectively. Between the encoder and decoder, high-resolution feature maps are bypassed to let the network also make use of local high-resolution information. The training was conducted using 80,000 iterations with a batch size of 32 samples. The batches were created by drawing random $128\times 128$ sub-crops from the training set. We used the Adam optimizer~\citep{kingma2014Adam} with a learning rate of 0.0001.

\subsubsection{Deep ensemble results}
\begin{figure}
    \centering
    \includegraphics[width=11.9cm]{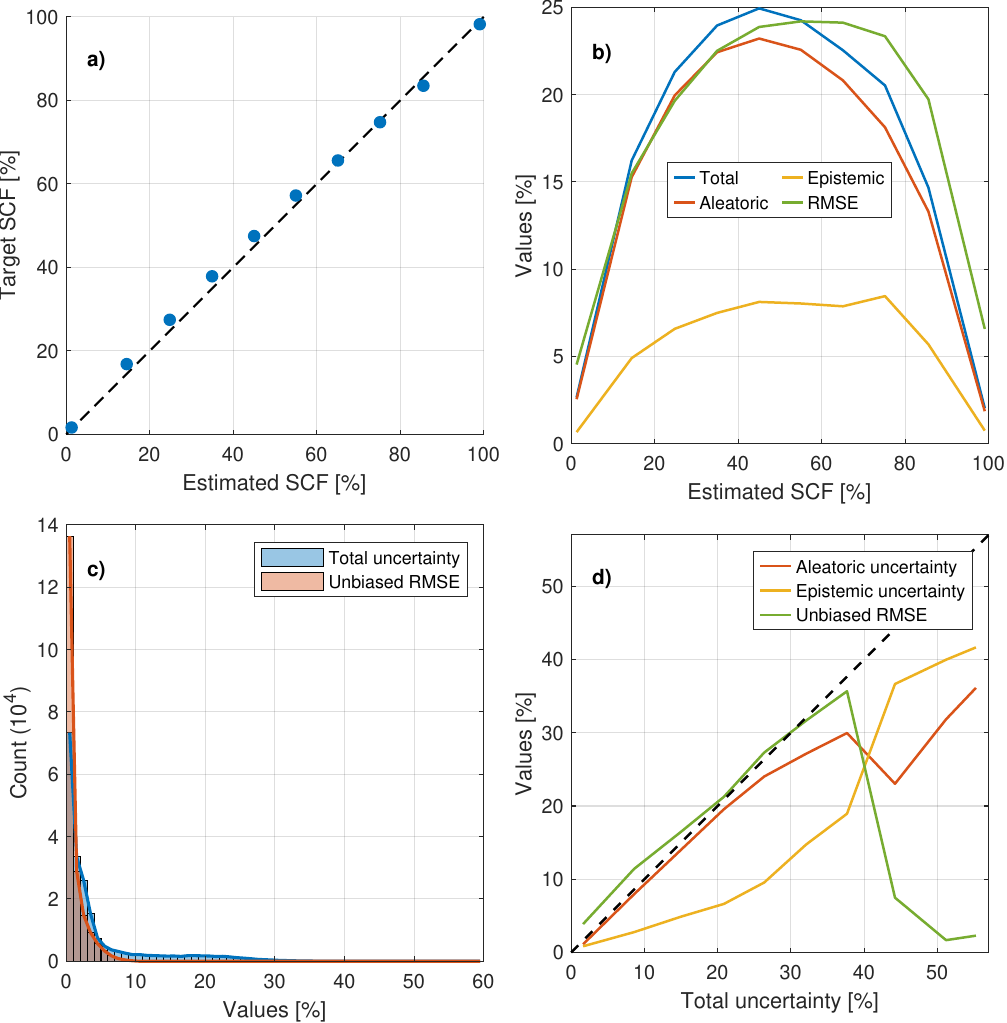}
    \caption{SCF estimations and their uncertainties. (a) Estimated SCF compared with ground truth. (b) Estimated SCF compared with epistemic, aleatoric and total uncertainty estimates and the unbiased RMSE. (c) Distribution of total uncertainty and unbiased RMSE. (d) Total uncertainty compared with aleatoric and epistemic uncertainties and unbiased RMSE.}
    \label{fig:results_DE_snow}
\end{figure}

A ten-member ensemble was created by training ten individual U-Net models to estimate the SCF values, together with the aleatoric and epistemic uncertainties as outlined in Section~\ref{sec:Deep ensembles}. The fidelity of these uncertainties was evaluated by categorizing the estimated SCF into 10 bins, calculating the average estimated SCF, and comparing it with the average target (ground truth) SCF for each bin using the test samples. The results indicate a strong correlation between the average estimated and target SCF values (Fig.~\ref{fig:results_DE_snow}a). Similar calculations applied to the uncertainty measurements reveal a close alignment between the total uncertainty (Eq.~\eqref{eq:DE_sig}) and the unbiased root mean squared error (RMSE) with respect to ground truth, albeit with a slight deviation in the unbiased RMSE compared to the total uncertainty (Fig.~\ref{fig:results_DE_snow}b). Further analysis shows that uncertainties are lowest at high and low SCF values, with aleatoric uncertainty consistently higher than the epistemic uncertainty. Distribution comparisons between the total uncertainty and the unbiased RMSE demonstrate that the former exhibited a heavier right tail, suggesting potential spikes in estimated uncertainties (Fig.~\ref{fig:results_DE_snow}c). To assess the performance further, we divide the estimated total uncertainty into 10 bins and compute the average aleatoric and epistemic uncertainties, and the average unbiased RMSE for each bin. The results indicate a very good correspondence between the average total uncertainty and the average unbiased RMSE for total uncertainty values less than \SI{38}{\percent} (Fig.~\ref{fig:results_DE_snow}d). However, for total uncertainty values exceeding 38, the unbiased RMSE decreases, with epistemic uncertainty becoming predominant at higher total uncertainty levels.

\begin{figure}
    \centering
    \mbox{}\hspace{1.5cm}
    \includegraphics[width=0.28\linewidth]{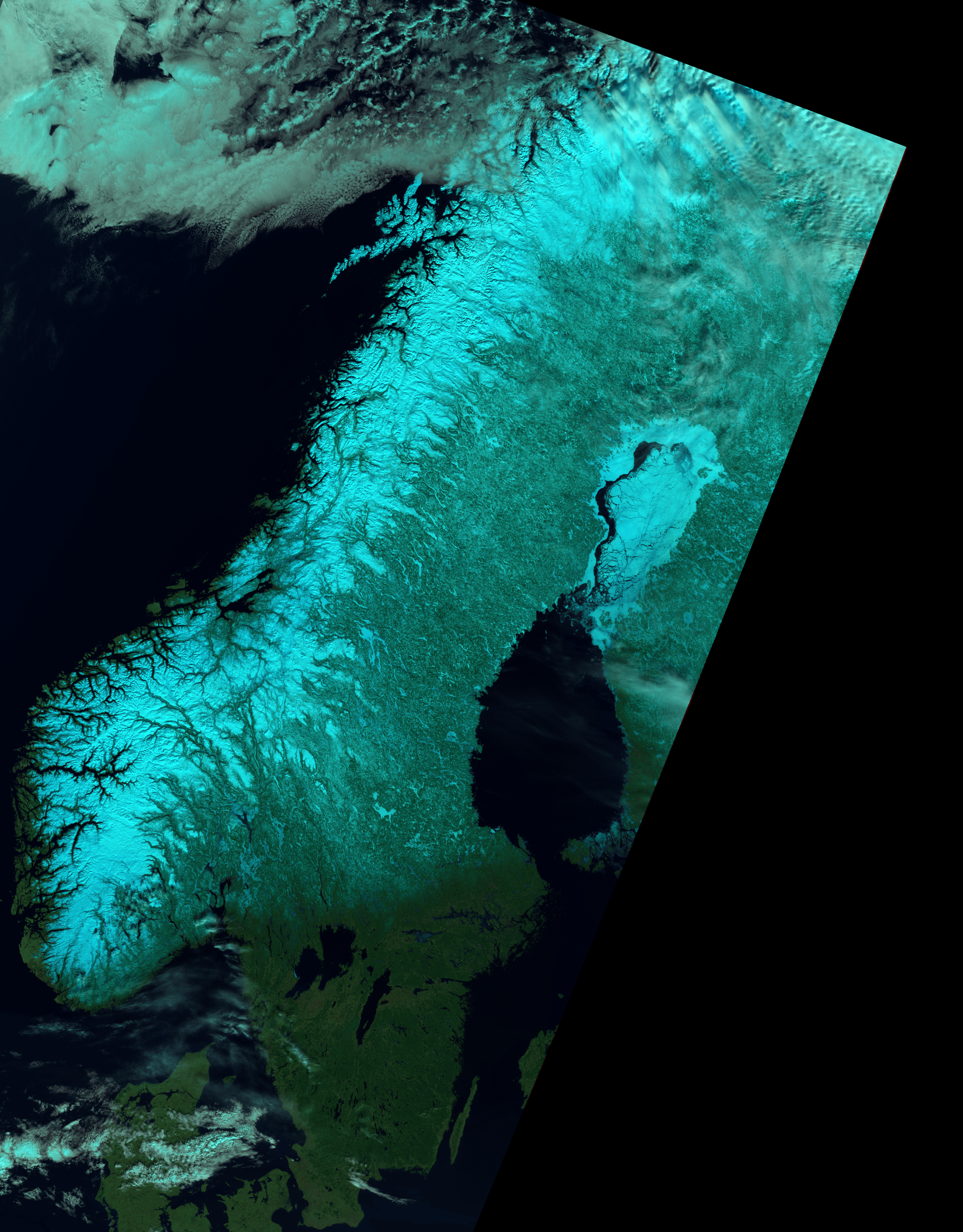}
    \includegraphics[width=0.28\linewidth]{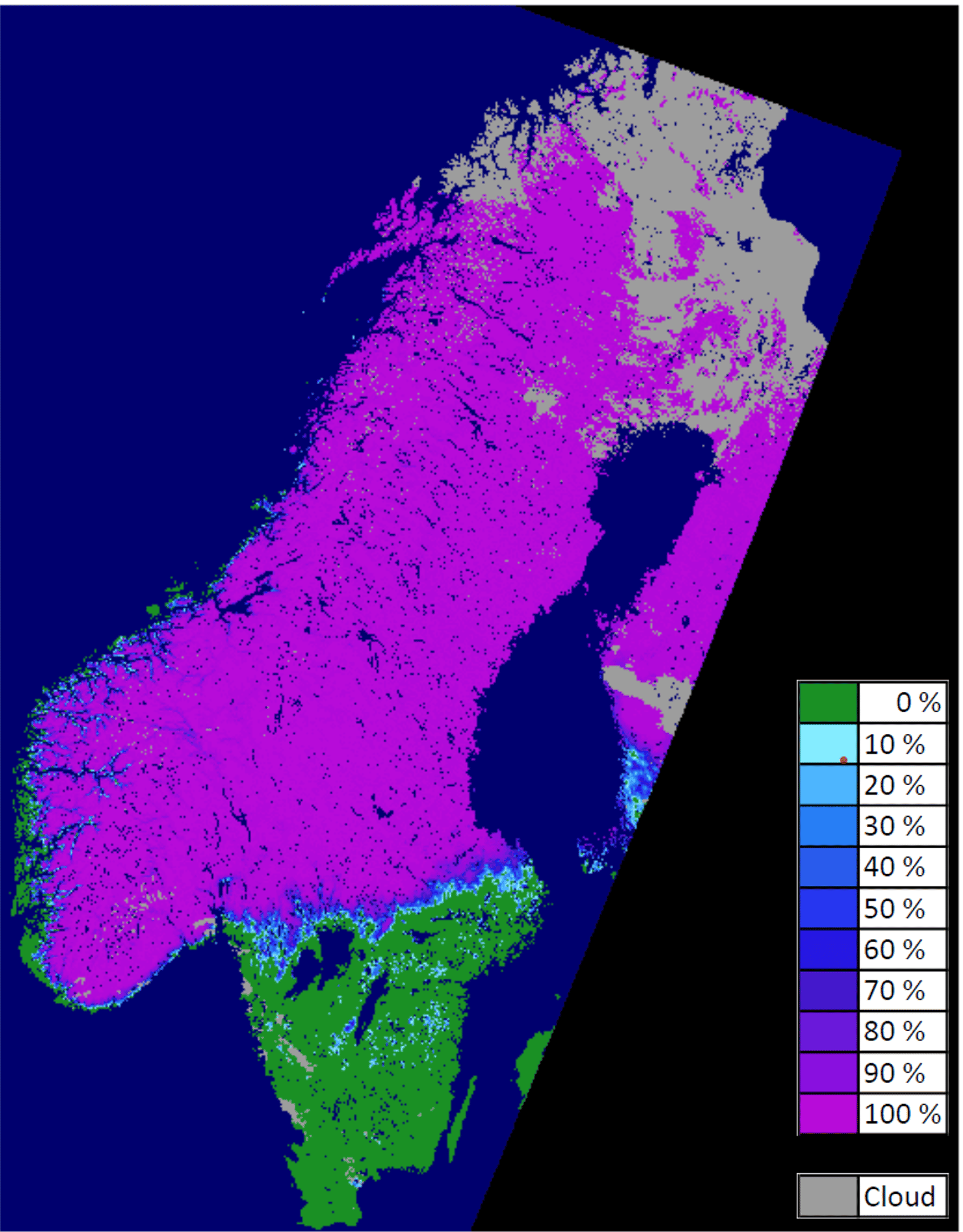}
    \newline
    \centering
    \includegraphics[width=0.28\linewidth]{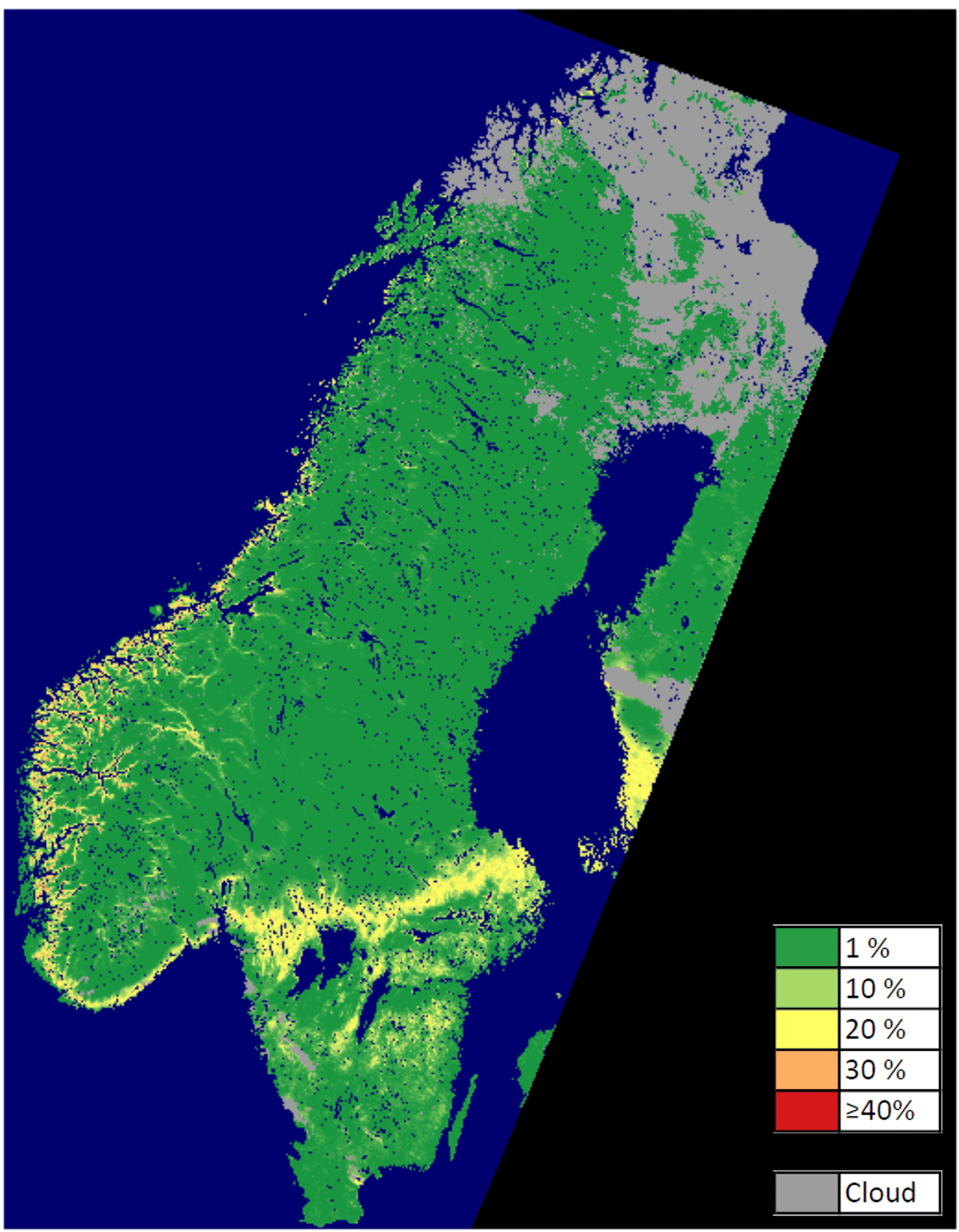}
    \includegraphics[width=0.28\linewidth]{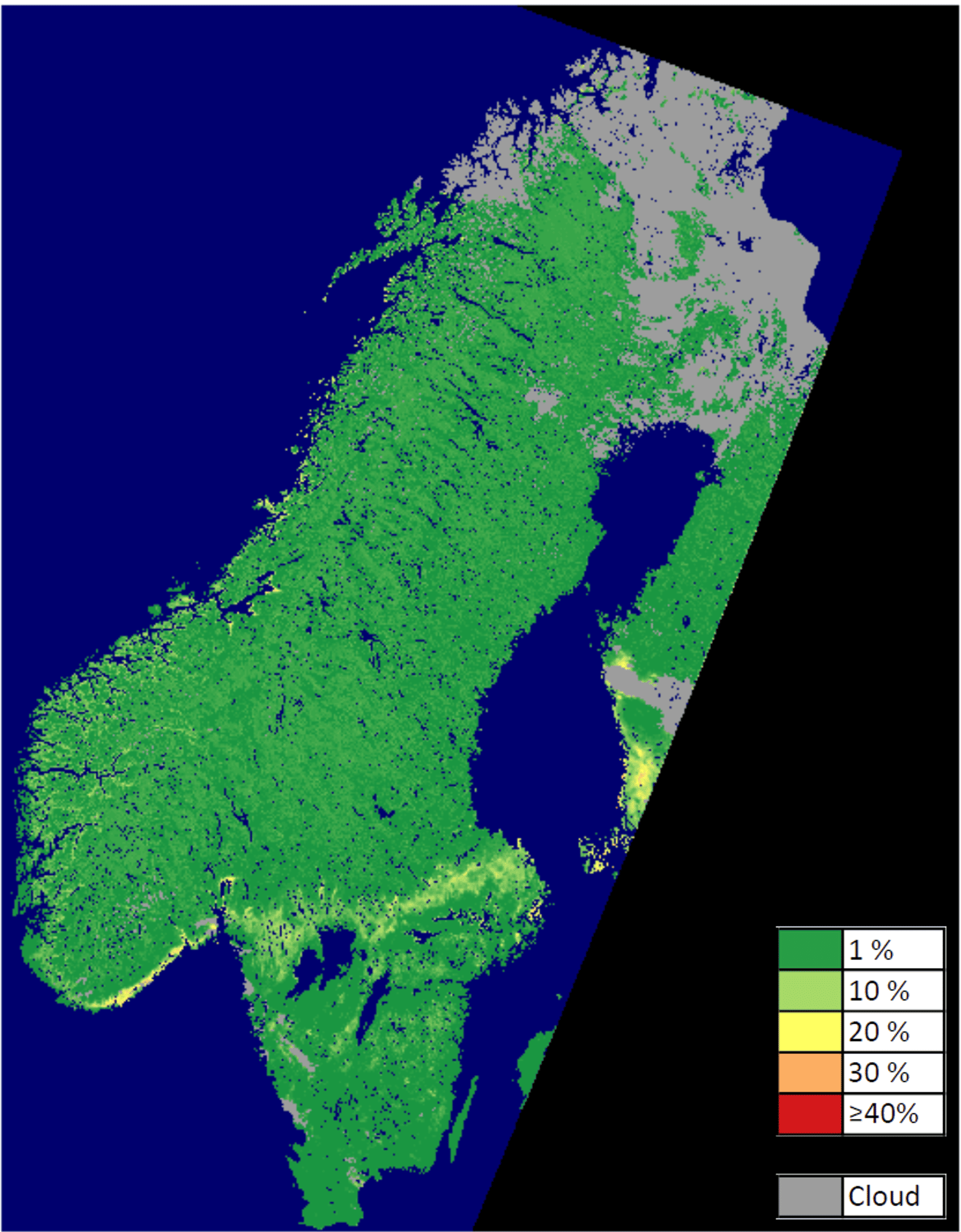}
    \includegraphics[width=0.28\linewidth]{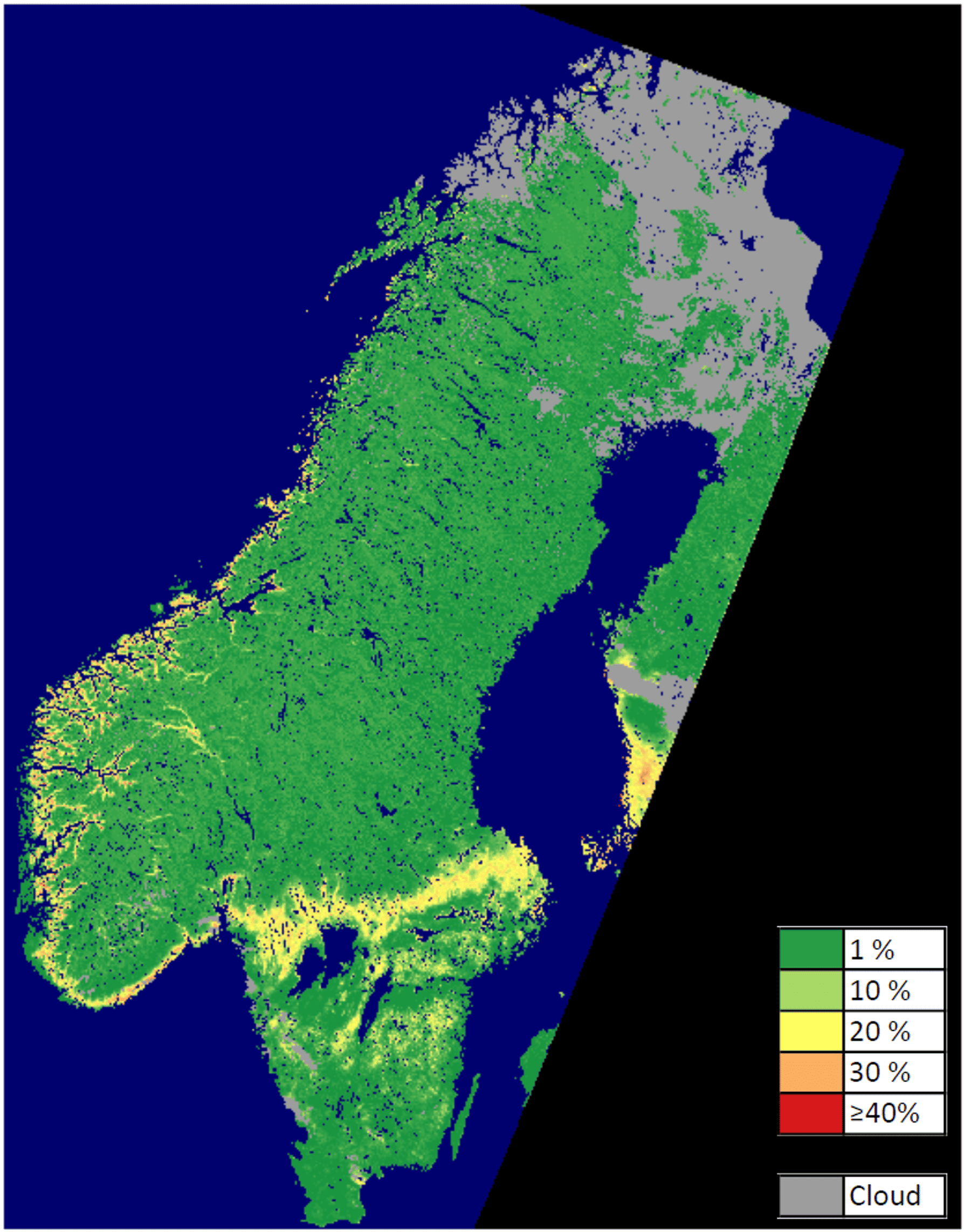}
    \caption{SCF estimation results and associated uncertainties obtained from deep ensemble for April 12, 2018. Upper-left: Sentinel-3 SLSTR data. Upper-right: Estimated SCF values. Lower-left: Aleatoric uncertainties. Lower-center: Epistemic uncertainties. Lower-right: Total uncertainties.}
    \label{fig:uncertainties_20180412}
\end{figure}
An example from April 12, 2018, a cloud-free day across much of the study area, reveals low aleatoric and epistemic uncertainties in regions with high estimated SCF, including forested areas (Fig.~\ref{fig:uncertainties_20180412}). The highest uncertainties are noted near the snow line, such as in southern Sweden. Across the study area, the aleatoric uncertainties are in general higher than the epistemic ones. Examination of the uncertainty maps in Fig.~\ref{fig:uncertainties_20180412} indicates similar spatial patterns in both types of uncertainty in the sense that the aleatoric uncertainty has high values at the same locations as the epistemic uncertainty has high values. This aligns with the findings in~\cite{kahl2024}, which demonstrated that, while separation of aleatoric and epistemic uncertainties is effective with simulated data, it may not directly translate to real-world scenarios. 

\begin{figure}
    \centering
    \mbox{}\hspace{1.5cm}
    \includegraphics[width=0.28\linewidth]{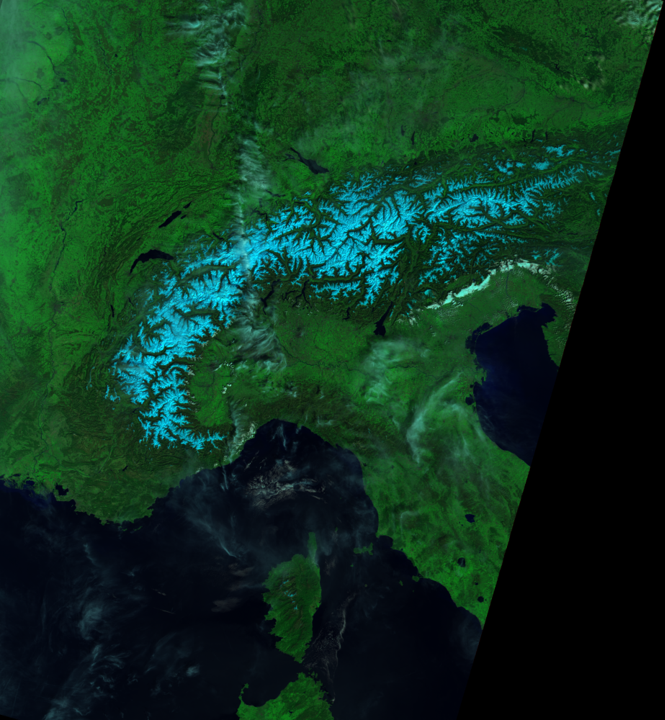}
    \includegraphics[width=0.28\linewidth]{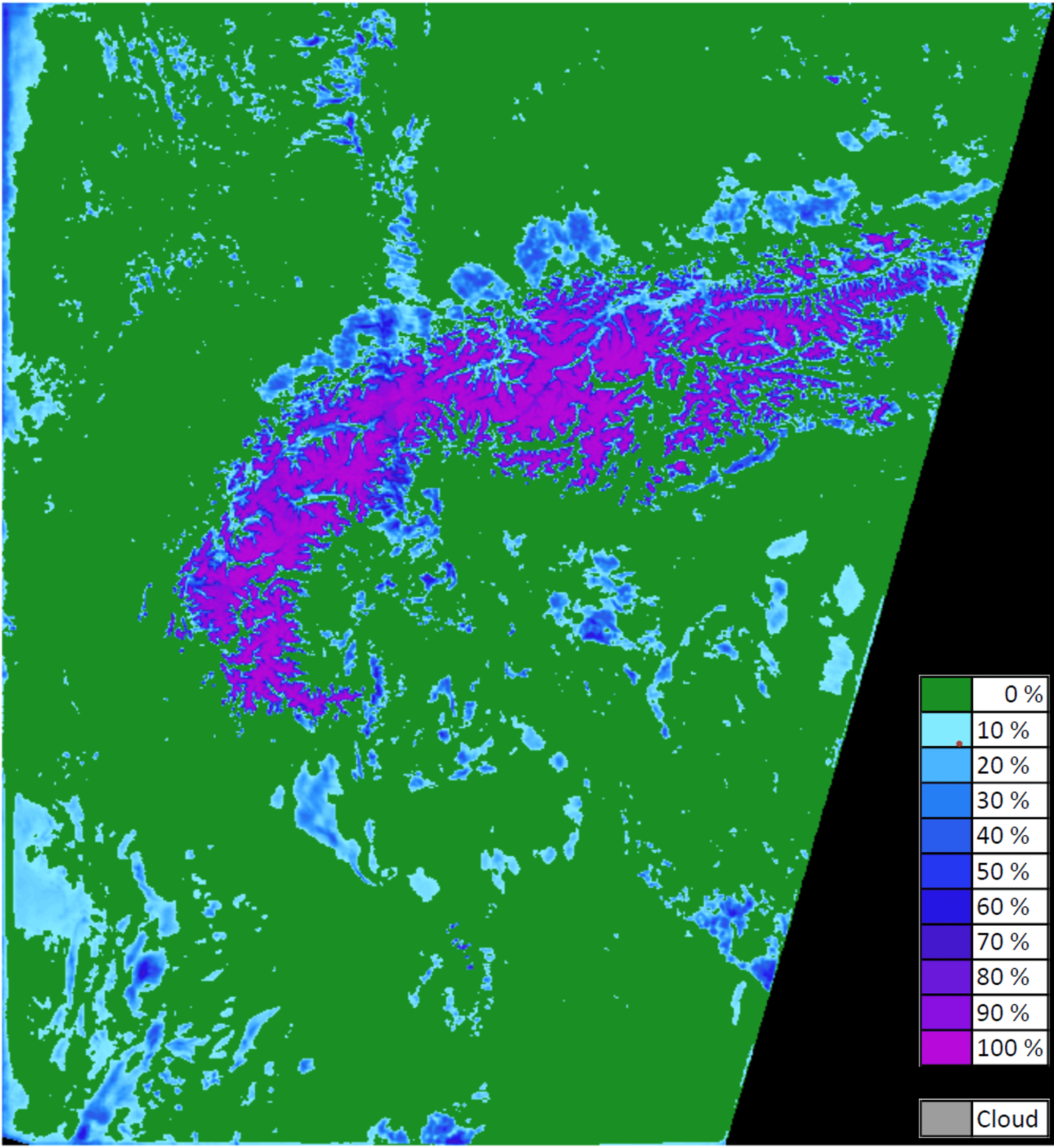}
    \newline
    \centering
    \includegraphics[width=0.28\linewidth]{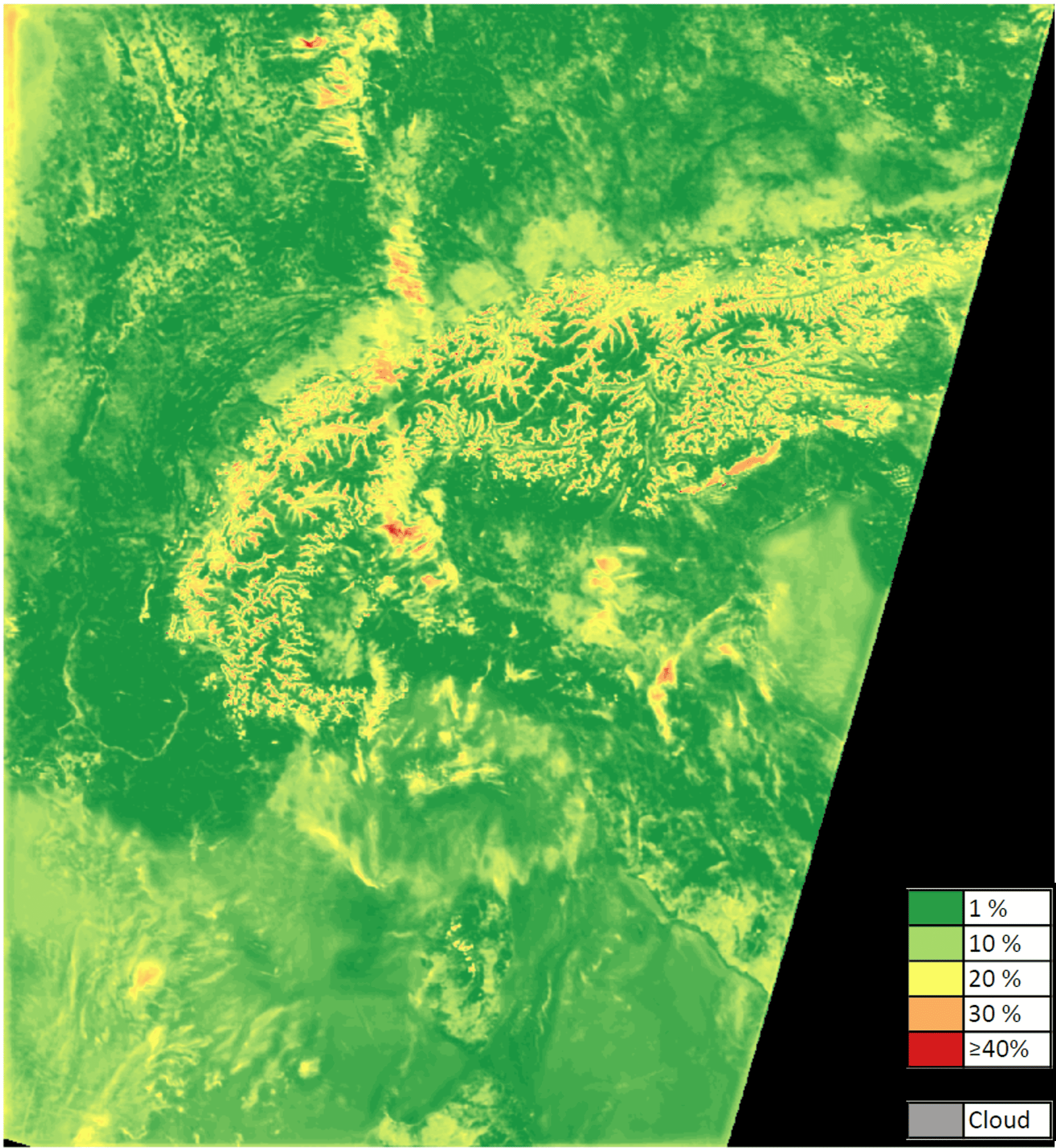}
    \includegraphics[width=0.28\linewidth]{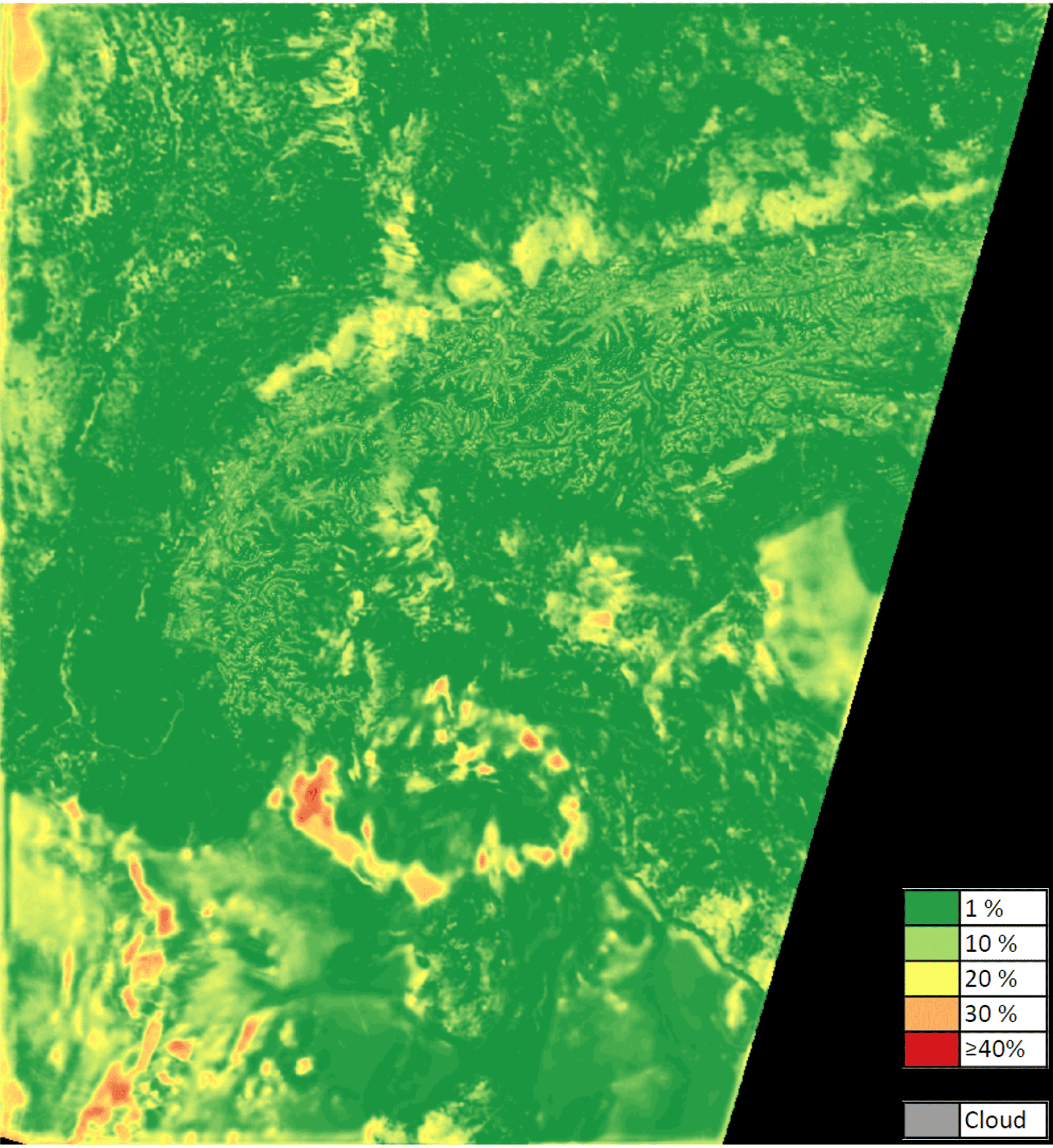}
    \includegraphics[width=0.28\linewidth]{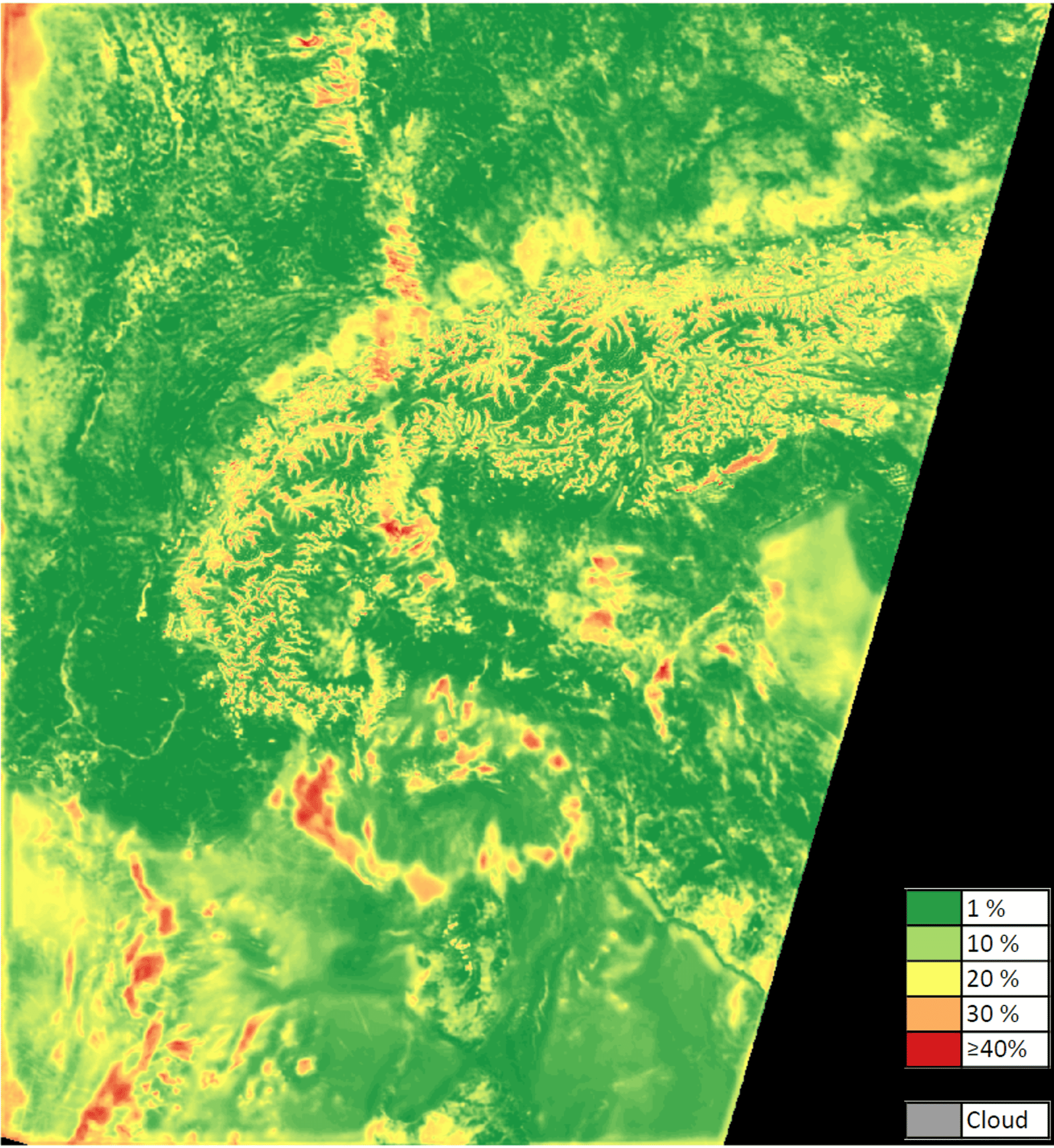}
    \caption{Results of deep ensemble uncertainties for the European Alps on April 6, 2024. Upper, left: Sentinel-3 SLSTR. Upper, right: Estimated snow cover fraction. Lower, left: Aleatoric uncertainty. lower, center: Epistemic uncertainty. Lower, right: Total uncertainty of SCF estimates.}
    \label{fig:uncertainties_alps_20240406}
\end{figure}

To further demonstrate uncertainty estimation using deep ensembles, we applied the deep learning model trained exclusively on Scandinavian data, to estimate SCF and corresponding uncertainties in the European Alps (Fig.~\ref{fig:uncertainties_alps_20240406}). Despite the lack of reference data for quality assessment and the absence of water and cloud masks, visual inspection suggests that the model accurately maps snow coverage in the Alps. However, it incorrectly predicts snow north of the Alps and misclassified clouds as snow (Fig.~\ref{fig:uncertainties_alps_20240406}, upper-right), likely due to the lack of cloud data in the training dataset and a domain shift in land cover and terrain between Scandinavia and the Alps. Yet, when inspecting the corresponding uncertainty maps, we observe that most of these erroneous classifications have high uncertainties. Similar to the Scandinavian test case, we observe that for lower SCF estimates, the aleatoric uncertainties are typically higher than the epistemic ones. However, for cloud objects we often observe higher epistemic uncertainties (Fig.~\ref{fig:uncertainties_alps_20240406}) with different patterns than the aleatoric uncertainties, contrary to the Scandinavian case. This phenomenon demonstrates that the main reason for epistemic uncertainties in a deep learning model is an incomplete coverage of the training data distribution. Tests on images with higher cloud coverage often result in thick cloud interiors being mistakenly estimated as \SI{100}{\percent} SCF, with near-zero corresponding uncertainties. Such scenarios have huge dataset shifts compared to the training data, and the findings support the conclusions of~\cite{ovadia2019} that the quality of the uncertainty products degrades with increasing dataset shift, despite the relative robustness of deep ensembles to such shifts.
\subsection{Terrestrial water storage changes}
\label{sec:Application-TWS}

\subsubsection{Overview of data and models}
\begin{figure}[!ht]
    \centering
    \includegraphics[width=11.9cm]{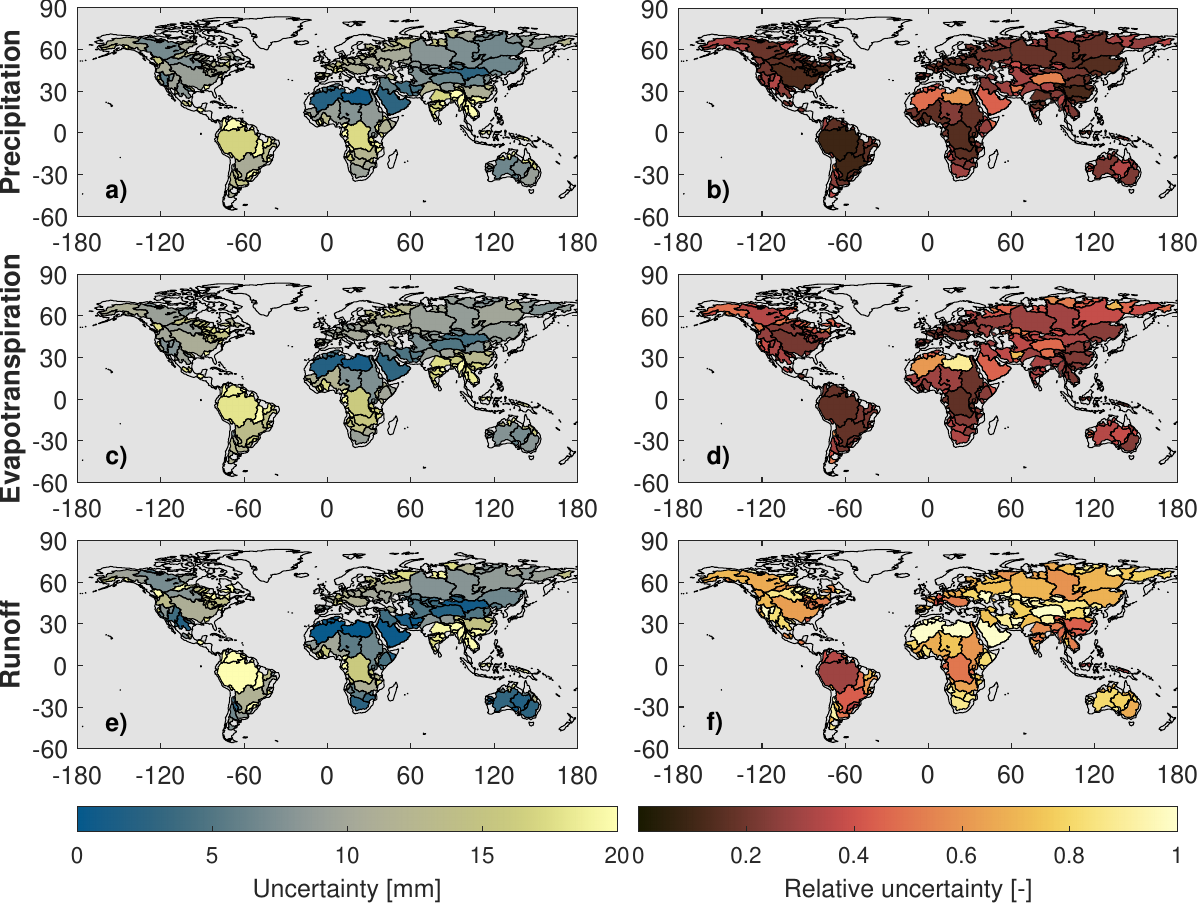}
    \caption{The uncertainties (left) and the ratio of uncertainties w.r.t. signal levels (right) of the three hydrological variables across multiple products.}
    \label{fig:Map_Data_Hydrology}
\end{figure}

In the second application case, we designed an experiment to predict the terrestrial water storage changes based on hydrological variables and focus on discussing the uncertainties quantified from different deep learning models. The problem setting is motivated by the water balance equation, which describes the relationship between the changes of TWS ($\mathrm{TWSC}$) and precipitation ($P$), evapotranspiration ($ET$) and runoff ($R$):
\begin{equation}
    \mathrm{TWSC} = P - ET - R,\label{eq:WaterBalance}
\end{equation}
where TWSC can be obtained from GRACE-measured TWS anomaly (TWSA) by computing centered finite differences:
\begin{equation}
    \mathrm{TWSC}_t = \frac{\mathrm{TWSA}_{t+1}-\mathrm{TWSA}_{t-1}}{2\Delta t},\label{eq:TWSC}
\end{equation}
with $t_1$, $t$, $t+1$ denoting three consecutive months, and $\Delta t$ denoting one month. The missing months within the study period were filled using cubic interpolation. The data were collected by~\cite{lehmann2022WBE}, including 11 sets of precipitation and runoff data and 14 sets of evapotranspiration data with different temporal coverage. Based on the temporal intersection of these datasets and the GRACE satellite mission, we selected data from January 2003 to December 2014. To reconcile the different spatial resolutions of the used datasets and suppress included outliers, basin-wise average time series in 189 river basins defined by the Global Runoff Data Center~\citep{GRDC2020} were generated. We used the ensemble average of all the mentioned datasets as the input and their standard deviations as associated uncertainty information for our deep learning approach. The standard deviations of individual products can usually result in relatively realistic uncertainty estimates but may also overestimate them due to the presence of outliers~\citep{tarek2021uncertainty,goswami2024water}. The average uncertainties of the three hydrological parameters of the 189 basins, both absolute and relative compared to the average signal levels in terms of standard deviations, are shown in Fig.~\ref{fig:Map_Data_Hydrology}.

Considering the high-frequency artifacts that may be introduced by simple forward or backward differences~\citep{landerer2010WBE} and different decaying of the daily fluxes in the storage~\citep{humphrey2016GRACETemporal}, we selected an input sequence of 11 months ($t-5,\dots,t,\dots,t+5$) to model the TWSC at the central month (epoch $t$) as the target. This problem setting allows the models to overcome the high-frequency artifacts based on optimization while considering the intra-annual signals, which should cover the varying water storage memories in most basins. To this end, we obtained 24'948 samples from January 2003 to December 2013 (132 months) and further split them into training set from January 2003 to December 2010 (96 months, 18'144 samples) and test set from January 2011 to December 2013 (36 months, 6'804 samples).

We selected four deep learning models, including MC-Dropout (Sec.~\ref{sec:MC-Dropout}), BNN (Sec.~\ref{sec:Bayes_by_Backprop}), deep ensemble (DE; Sec~\ref{sec:Deep ensembles}) and combining deep ensemble and Monte Carlo simulations (DEMC; Sec.~\ref{sec:DEMC}). We further applied the least-square adjustment to estimate the linear relationship between the 33 input features and the one target to show the quantified uncertainties in a conventional statistical model. The comparison focuses on the characteristics of obtained uncertainties using different techniques but refrain from evaluating the performance in terms of accuracy. The model architectures of different models and the optimizing strategies stay the same to ensure a relatively fair comparison among the different candidates. We designed an MLP model with five layers. The first layer (input layer) projects 33 features into a 256-dimensional hidden state and feeds to three hidden layers with 256 neurons. The last layer (output layer) projects the final hidden state into two outputs, representing the targeted TWSC at epoch $t$ and its associated aleatoric uncertainty. The final layer of the DEMC model is slightly different since it only provides one output representing the targeted TWSC at epoch $t$ (Sec.~\ref{sec:DEMC}). All the layers except for the output layer are followed by a $\mathrm{ReLU}(x)=\max(x, 0)$ activation function. All the models were implemented in PyTorch V2.1.0~\citep{paszke2019pytorch} and optimized using Adam~\citep{kingma2014Adam}. Specifically, the BNN model has been realized using a publicly available implementation by \cite{lee2022graddiv}.

\subsubsection{Comparing the quantified uncertainties using different methods}
\begin{figure}[!ht]
    \centering
    \includegraphics[width=11.9cm]{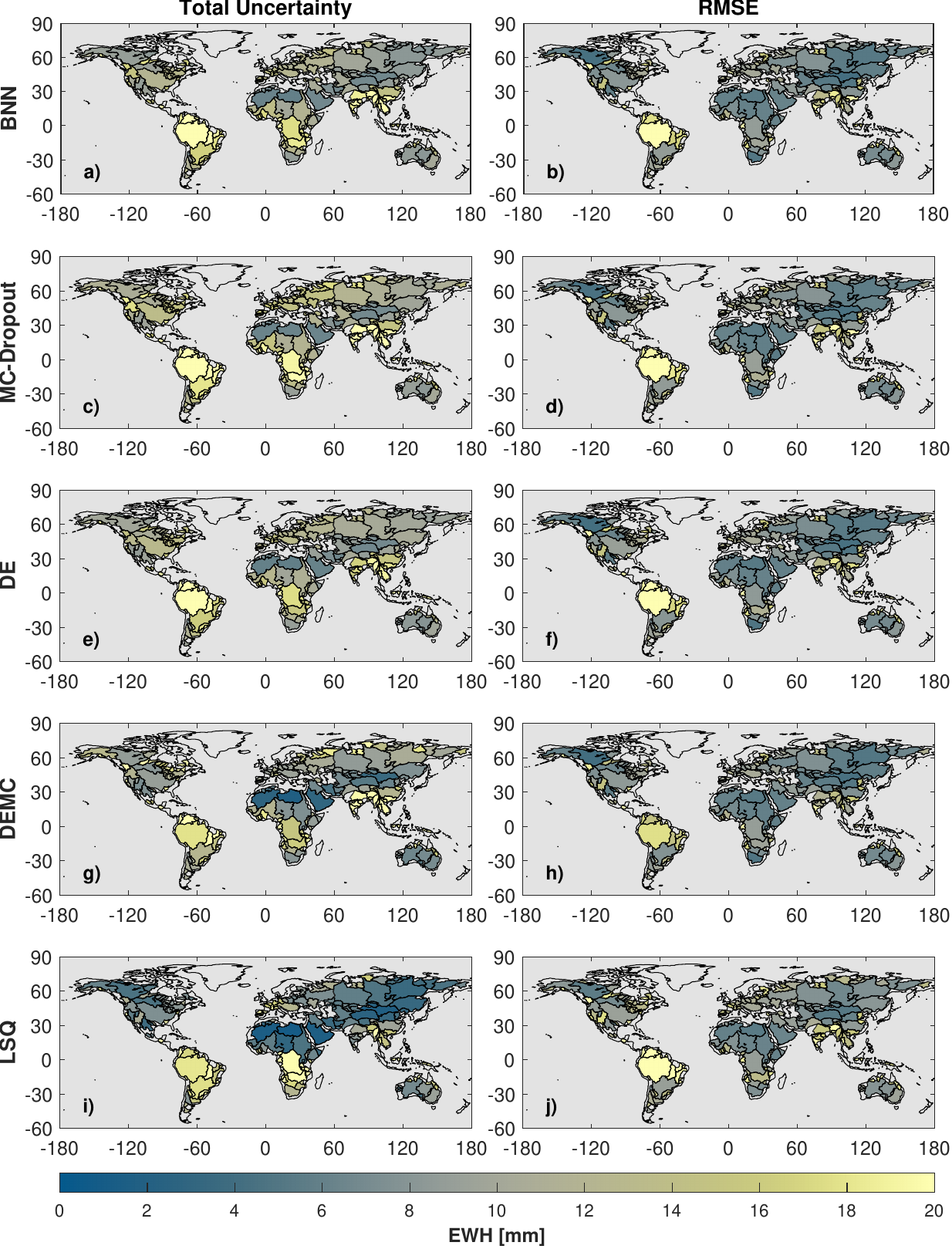}
    \caption{Comparison of the total predictive uncertainties (left) and errors of the predictions (right) in the 189 basins. The five rows show the results using four deep learning models and least-square estimation (LSQ). All the values are in the format of equivalent water height (EWH) with millimeters as units.}
    \label{fig:Map_Compare_TotalUncertainty}
\end{figure}

Fig.~\ref{fig:Map_Compare_TotalUncertainty} shows the average total predictive uncertainties and prediction errors in terms of RMSE over the three-year test interval. All the deep learning methods provide plausible uncertainties with BNN and MC-Dropout tending to provide relatively higher uncertainty estimates than DE and DEMC. The uncertainties estimated by the least-square model are slightly lower than others and tend to be over-optimistic (Fig.~\ref{fig:Map_Compare_TotalUncertainty}i and j). The unconsidered epistemic uncertainties likely cause this underestimation, as discussed in Section~\ref{sec:Statistical view}.

We further analyze the relationship between prediction errors and the predictive uncertainties of individual samples, namely individual months in all 189 basins. Table~\ref{table:Confidence_Inverval} shows the percentage of prediction errors lying within the corresponding confidence interval. All the deep learning methods have similar confidence intervals, with MC-Dropout tending to overshoot the prediction errors, whereas the least-square model clearly underestimates the total uncertainties. The deep learning methods typically have a wider 1-$\sigma$ confidence interval but a narrower 3-$\sigma$ interval compared to a Gaussian distribution, indicating that the distribution of errors is not fully Gaussian but with long-tail outliers, which cannot be avoided in real-world applications.

\begin{table}[!ht]
    \caption{The percentages of prediction errors that lie in the confidence intervals for the five models considering their quantified uncertainties. The reference confidence intervals by assuming a Gaussian distribution are \SI{68}{\percent}, \SI{95}{\percent}, and \SI{99.7}{\percent}, for one, two, and three standard deviations, respectively.}
    \centering
    \begin{tabular}{c|ccc|ccc}
        \toprule
        & \multicolumn{3}{c|}{Training set}& \multicolumn{3}{c}{Test set}\\
        & 1-$\sigma$ & 2-$\sigma$& 3-$\sigma$ & 1-$\sigma$& 2-$\sigma$& 3-$\sigma$\\
        \midrule
        BNN& \SI{76.1}{\percent}& \SI{96.8}{\percent}& \SI{99.5}{\percent}& \SI{67.7}{\percent}&\SI{93.1}{\percent}& \SI{98.8}{\percent}\\
        MC-Dropout& \SI{79.6}{\percent}& \SI{97.4}{\percent}& \SI{99.6}{\percent}& \SI{72.1}{\percent}&\SI{95.1}{\percent}& \SI{99.2}{\percent}\\
         DE& \SI{75.3}{\percent}& \SI{96.2}{\percent}& \SI{99.3}{\percent}& \SI{67.3}{\percent}&\SI{93.0}{\percent}& \SI{98.6}{\percent}\\
        DEMC& \SI{75.8}{\percent}& \SI{94.7}{\percent}& \SI{98.5}{\percent}& \SI{66.7}{\percent}& \SI{91.2}{\percent}& \SI{97.6}{\percent}\\
        LSQ& \SI{66.4}{\percent}& \SI{89.7}{\percent}& \SI{96.3}{\percent}& \SI{61.7}{\percent}& \SI{88.1}{\percent}& \SI{95.9}{\percent}\\
        \bottomrule
    \end{tabular}
    \label{table:Confidence_Inverval}
\end{table}

\begin{figure}[!ht]
    \centering
    \includegraphics[width=11.9cm]{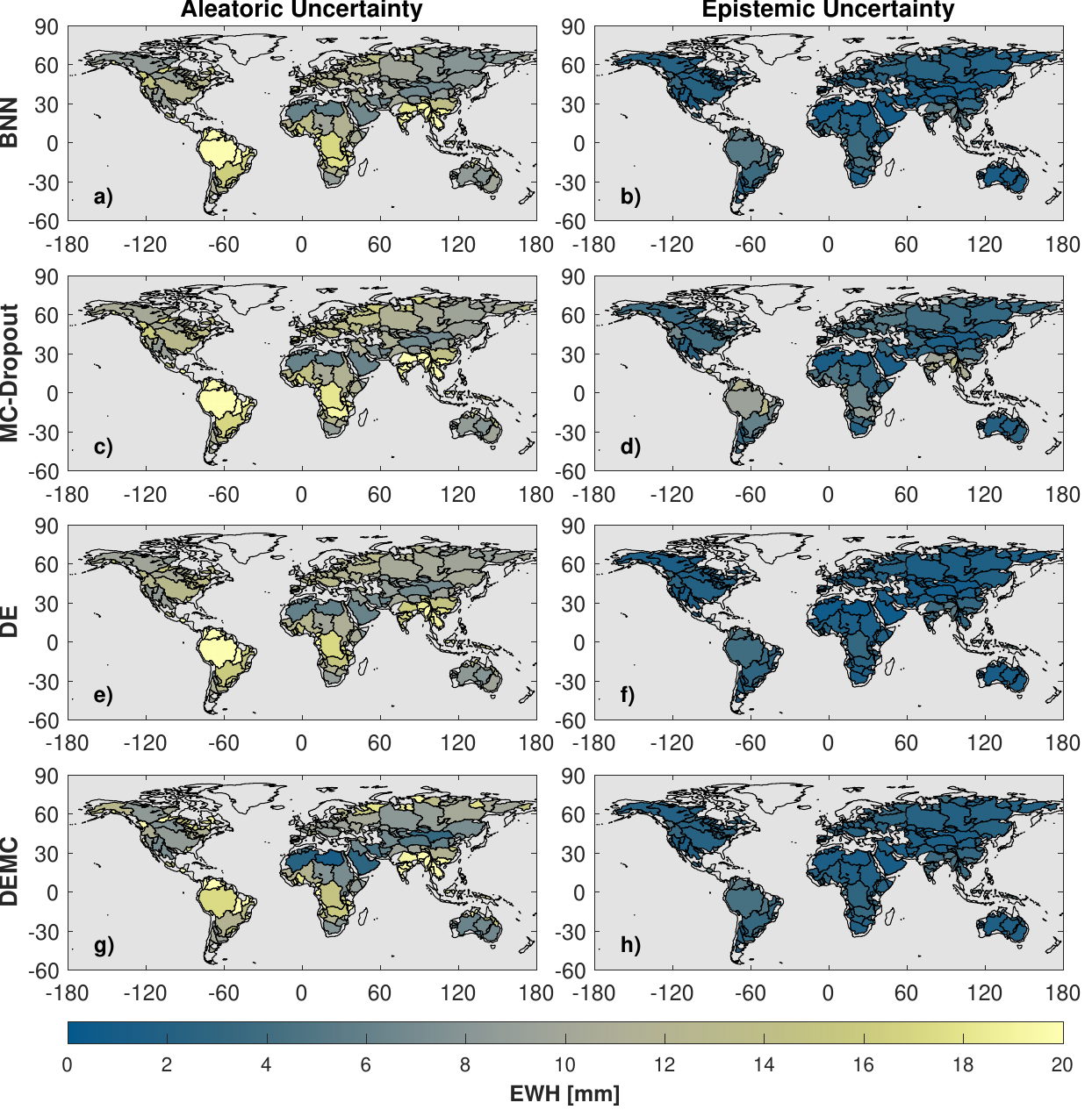}
    \caption{Comparison of the aleatoric uncertainties (left) and epistemic uncertainties (right) in the 189 basins. The four rows show the results using four different deep learning models. All the values are in the format of equivalent water height (EWH) with millimeters as units.}
    \label{fig:Map_Compare_AleatoricAndEpistemic}
\end{figure}

Fig.~\ref{fig:Map_Compare_AleatoricAndEpistemic} further extends the analysis by showing the aleatoric and epistemic uncertainties separately. It is worth noting that the aleatoric uncertainties provided by BNN, MC-Dropout, and DE models are directly generated by the network based on Eqs.~\eqref{eq:Distribution_NLL} to \eqref{eq:Loss_NLL}, while the ones from DEMC were generated based on sampling considering the input uncertainties, see Eqs.~\eqref{eq:MCDE_mu} and \eqref{eq:MCDE_sig}. Aleatoric uncertainties with marginal contributions from the epistemic uncertainties dominate the total predictive uncertainties of all four tested deep learning models. The results prove the efficiency of the tested non-BNN proxies since they provide epistemic uncertainties that are close to those of the BNN ones. Moreover, the low epistemic uncertainties suggest that the training set has relatively good coverage of the whole data distribution and, therefore, the model has good generalizability. Only MC-Dropout provides slightly higher epistemic uncertainties, which are due to the varying model architecture as described in Section~\ref{sec:MC-Dropout}. The dominating role of aleatoric uncertainties in this application case is similar to the previous snow example in the Scandinavian Peninsula and opposite to the European Alps case. This application case indicates the potential risk of heavily underestimating the quantified total predictive uncertainties if the aleatoric uncertainties are not rigorously considered.

Also noteworthy are the different patterns of aleatoric uncertainties. Aleatoric uncertainties estimated by the DEMC model show slightly different patterns than the other three models. Higher aleatoric uncertainties are reported by DEMC for the basins in North America, including the Fraser Basin, Churchill Basin, and other small basins surrounding Hudson Bay, as well as for the basins in Northern Siberia. This effect results from higher runoff uncertainties in those basins (Fig.~\ref{fig:Map_Data_Hydrology}e). The results demonstrate the ability of DEMC to provide realistic uncertainty information by explicitly considering input uncertainties, whereas the other methods tend to underestimate the total uncertainties in these regions (Fig.~\ref{fig:Map_Compare_TotalUncertainty}). Therefore, DEMC can alert potential inferior estimations when the inputs are associated with high uncertainties, which is valuable for satellite-based variable parameter estimations, as long as we have access to input uncertainties.
\section{Conclusions and Future Perspectives}
\label{sec:Conclusions and Future Perspectives}

\subsection{Conclusions}
Reliable uncertainty information is critical for decision-making, adding value to results and making them trustworthy. This survey starts with the theoretical foundations of uncertainties to discuss the similarities and differences between conventional statistical and deep learning perspectives on problem setting and uncertainty quantification. The main takeaway is that the conventional statistical view assumes that the predefined relationship based on domain knowledge is correct, so all the uncertainty should stem from input data randomness. Although the deep learning perspective also presumes the existence of a relationship between inputs and targets, the functional expression of such a relationship is considered unknown and empirically estimated from large training datasets using a model with sufficient capacity. Deep learning users aim to understand the deficiency of the models (epistemic uncertainties) and allow the model to quantify the data uncertainties (aleatoric uncertainties) based on the overall data distribution, while typically ignoring the quality of individual samples. Both perspectives are based on the different nature of the respective problem settings. Applying deep learning algorithms to sEO data for ECV estimation is an interdisciplinary problem intersecting classical Earth science and advanced data-driven methods. Hence, we need to bridge the different perspectives and combine them to consider their individual strengths and limitations so that we can fulfill the requirements from both sides. The geoscience and climate scientists can benefit from the matured uncertainty quantification approaches developed for typical deep learning tasks such as vision and language modeling~\citep{abdar2021UQinDLsurvey,gawlikowski2023UQinDLsurvey}, with some specific modifications aimed at solving specific ECV-related problems.

In Section~\ref{sec:UQ-DL}, we provide a thorough discussion on the different properties of various deep learning methods with examples from previous ECV studies. This discussion provides a basis for selecting suitable deep learning methods for the specific requirements of a certain task. Furthermore, this study applied selected BDL techniques to two real-world cases of ECV retrieval to provide quantitative assessments of uncertainties obtained from different approaches and demonstrates that the different BDL methods work equally well with marginal differences among them. However, this may be a partial conclusion since we only tested two cases, which do not represent the entire Earth system. An important conclusion is that we cannot neglect either aleatoric or epistemic uncertainties since the dominating role varies across different applications, and even specific cases. It is highly recommended that both types of uncertainties be rigorously quantified to provide realistic total uncertainties unless there is strong evidence that one of them is negligible. Explicitly considering the input feature uncertainties when quantifying the predictive uncertainties has proven beneficial for ECV applications. Whenever such information is available, it should thus be considered. On the other side, the conventional least-square adjustment based on predefined functional relationships clearly underestimates the predictive uncertainties in our application case, highlighting the impact of ignoring epistemic uncertainties.

\subsection{Future perspectives}
One of the major concerns about using sEO products derived from deep learning techniques is the "black-box" nature of deep learning algorithms. These characteristics impede users from understanding the model, which can lead to doubts about the results, especially when making sensitive decisions. Realistic uncertainty quantification enhances trust in the results and, furthermore, can be seen as a contribution to explainable AI~\citep{roscher2020xML,samek2021xDL}. Interpretable machine learning has been gaining increasing attention in the field of geoscience to avoid potential risks caused by superficial applications and better utilize the potential of machine learning~\citep{jiang2024IML4Geoscience,kiani2024xAI}.

Another concern that needs to be addressed is the appropriate consideration of input data uncertainties. It is overly optimistic to expect a world where all sources of uncertainty are known. However, uncertainty information is more likely to be available for sEO datasets than for data types typically used in deep learning tasks, such as texts and photos. Therefore, explicitly considering the input uncertainties for both ECV retrieval and associated uncertainty quantification should be promoted. Geoscientists and climate scientists may take the responsibility to develop suitable tools for this purpose since this direction may not fully align with the priorities for vision and language modeling applications, in which the data uncertainties are typically inaccessible.

Another important development is to combine physical information and data-driven approaches, for example, in the form of physics-informed neural networks (PINNs; \citealt{raissi2019PINNs}) or physics-constraint neural networks. When we have strong confidence in physics-based relationships and believe they are suitable for the specific application, we can impose this prior knowledge to enhance the performance of models~\citep[e.g.,][]{kiani2024PINNs}. We can either apply the constraints to the model parameters (the principle of PINNs) or the model predictions—the latter shares similar logic with the condition matrix $\mathbf{B}$ in a least-square adjustment. Alternatively, we can formulate the loss function based on such a relationship and convert a regression problem into a parameter estimation problem~\citep{kutz2023ML4PE}. To this end, we constrain the model parameters so that they are not entirely free to take arbitrary values from the parameter space but are forced to converge to the solutions fulfilling the physical relationships. With these developments, we may reduce the uncertainty by benefiting from the well-established relationship and also possibly better understanding the source of uncertainties. Conversely, deep learning algorithms may enable a more efficient way for scientific discovery by extracting valuable information from data and enhancing our domain expertise~\citep{wang2023AI4Science}. From this perspective, deep learning and physics-based methods are not in conflict but represent a complementary approach to be jointly advanced~\citep{levine2024MLandPhysics}, also in view of better quantifying and understanding the involved uncertainties.

\section*{Acronyms}
\begin{acronym}
\acro{AI} Artificial Intelligence
\acro{ATBD} Algorithm Theoretical Basis Document
\acro{BCNN} Bayesian Convolutional Neural Network
\acro{BDL} Bayesian Deep Learning
\acro{BLR} Bayesian Linear Regression
\acro{BNN} Bayesian Neural Network
\acro{C3S} Copernicus Climate Change Service
\acro{CNN} Convolution Neural Network
\acro{CLMS} Copernicus Land Monitoring Service
\acro{DE} Deep Ensembles
\acro{ECV} Essential Climate Variable
\acro{ELBO} Evidence Lower Bound
\acro{EO} Earth Observation
\acro{ESA CCI} European Space Agency Climate Change Initiative
\acro{EWH} Equivalent Water Height
\acro{GCOS} Global Climate Observing System
\acro{GRACE} Gravity Recovery and Climate Experiment
\acro{GRACE-FO} Gravity Recovery and Climate Experiment Follow-On
\acro{HMC} Hamiltonian Monte Carlo
\acro{KL} Kullback-Leible
\acro{LoRA} Low-rank Adaptations
\acro{LSTM} Long Short-Term Memory
\acro{MC} Monte Carlo
\acro{MC-Dropout} Monte Carlo Dropout
\acro{MCMC} Markov Chain Monte Carlo
\acro{MLE} Maximum Likelihood
Estimation
\acro{MLP} Multilayer Perception
\acro{NOAA} National Oceanic and Atmospheric Administration
\acro{PINN} Physics-informed Neural Network
\acro{PODAAC} Physical Oceanography Distributed Active Archive Center
\acro{RMSE} Root Mean Squared Error
\acro{SCF} Snow Cover Fraction
\acro{SGD} Stochastic Gradient Descent
\acro{SWA} Stochastic Weight Averaging
\acro{SWAG} Stochastic Weight Averaging with a Gaussian
\acro{sEO} Satellite Earth Observation
\acro{TWS} Terrestrial Water Storage
\acro{TWSA} Terrestrial Water Storage Anomaly
\acro{TWSC} Terrestrial Water Storage Change
\acro{VI} Variational Inference
\acro{xAI} Explainable Artificial Intelligence
\end{acronym}
  
\newpage

\backmatter

\section*{Declarations}
\bmhead{Acknowledgments}
This paper is an outcome of the Workshop “Remote Sensing in Climatology: Essential Climate Variables and their Uncertainties” held at the International Space Science Institute (ISSI) in Bern, Switzerland, 13-17 November 2023.

\bmhead{Author contributions}
Conceptualization: JG, AS, BS. Data curation \& Investigation: JG, AS, BS, AUW. Visualization: JG, AS. Discussion - ECV and conventional methods: JG, MJT, UM, EB, IV, AJ. Discussion - Deep learning: JG, AS, MKS, FD, KS, BS. Writing – initial draft: JG, AS, MKS, MJT, UM, EB, IV. Writing – review \& editing: All authors.

\bmhead{Funding}
AS, AUW, and FD were supported by the KnowEarth project, Norwegian Research Council (grant 337481) and the ESA project AI4Arctic.

\bmhead{Competing interests}
The authors declare no competing interests.

\bmhead{Data availability}
The used datasets are declared in the text with proper references. All the other intermediate data generated during this study are available from the corresponding author upon a reasonable request.

\bmhead{Code availability}
The used open-source codes have been declared in the text. The original codes generated during this study are available from the corresponding author upon a reasonable request.

\clearpage
\begin{appendices}
\section{Processing pipeline of GRACE(-FO) data and associated uncertainty sources}
\label{appendix:Processing pipeline of GRACE(-FO) data and associated uncertainty sources}
To explain the sources of uncertainties in a typical satellite-based parameter estimation pipeline in more details as an addition to Section~\ref{sec:Data processing pipeline and sources of uncertainties}, we further describe the multiple levels of products and their associated uncertainty propagation in a processing pipeline using the GRACE(-FO) satellite missions as an example.

Temporal variations in the mass distribution at the Earth's surface are derived from their impacts on satellite trajectories. Therefore, one of the main observable quantities is the satellites' absolute positions, which are determined in a precise orbit determination (POD) with centimeter accuracy from GPS code and phase observations~\citep{jaeggi2006}. In the case of the dedicated gravimetry missions GRACE~\citep{Tapley2004} and GRACE-FO~\citep{landerer2020}, additional K-band and laser range (only for GRACE-FO) observations of the inter-satellite range variations of the two satellites, can be obtained with sub-micrometer per second accuracy. The inter-satellite range observations are differentiated into range rates to reduce the impact of low-frequency errors. In both cases, the GPS phase and the K-band microwave observations, the integer phase ambiguities are not observable. The processing from raw measurements (L0) to preprocessed observations (L1) is usually done by the space agency (here, JPL) and may include irreversible steps~\citep{Wen2019GRACEL1handbook} and release to public for further processing. At this stage, the preliminary errors come from the inherent randomness in the measuring sensors (preliminary for L0) but also include epistemic uncertainties that come from potential model deficiencies (preliminary for L1), see Fig.~\ref{fig:Estimating_pipeline}.

The direct interpretation of L1 observations in the sense of gravity variations or mass changes is not possible and requires further complex inversion processing steps, known as L1 to L2. The data processing from L1 products to L2 products include retrieving geophysical signals from the observed data~\citep{mittaz2019applying}. In the case of the GRACE(-FO) missions, we need additional information to separate the impacts of the gravitational forces on the orbits of satellites from the impacts of the non-gravitational forces, such as atmospheric drag, solar radiation pressure, and Earth re-radiation. For example, the surface forces acting on the GRACE(-FO) satellites are measured by accelerometers located in the center of mass of the individual satellites, and the satellites' attitudes are determined from star camera observations. All these observations come with observation noises, the characteristics of which is rather poorly known. Moreover, the systemic problems of the accelerometers that occurred at the final stage of GRACE, and the beginning of GRACE-FO missions also brought considerable observation uncertainties. Once the surface forces have been removed from the observations, the remaining gravitational force has to be further reduced because many mass transport processes in the Earth system happen at frequencies that are too high to be resolved by the snapshot solutions of the gravity field, which are derived at monthly intervals. This signal separation step is known as de-aliasing. It is based on models of the mass effects of the ocean, solid Earth, atmosphere tides, and the non-tidal atmosphere and ocean variations, caused mainly by weather phenomena. Again, the tide models and the non-tidal atmosphere and ocean de-aliasing model (AOD; \citealt{shihora2022AOD1BRL07}) rely on observations and come with error budgets, which are not sufficiently known.

The L2 products can be further processed to generate L3 or L4 products, which are typically gridded and may include information from other geophysical data. A typical L3 product from GRACE(-FO) satellites is the gridded mass anomaly product in terms of equivalent water height~\citep{Boergens2021b,Dahle2024}, whereas L4 products are further processed to generate high-level variables like global mean sea level changes or environmental extremes indicators~\citep{tapley2019GRACE4Climate}. In the processing from L2 to L3, additional geophysical background models, such as those for glacial isostatic adjustment or earthquakes, are usually applied, introducing further uncertainties~\citep{Dahle2024}. Due to these diverse and poorly known sources of uncertainty, formal uncertainty propagation on the gravity field coefficients does not provide realistic results~\citep{Boergens2022}. The individual analysis centers of GRACE(-FO) data developed individual approaches to absorb noise by dedicated parameters~\citep{beutler2010a,beutler2010b} or to consider it by empirical noise models~\citep{kvas2019}. Consequently, the uncertainty information provided with the gravity field solutions is also very diverse, and again, propagation of the uncertainties to mass transport products of interest for the users of the ECVs is currently not feasible.

\section{Sensitivity Analysis}
\label{appendix:Sensitivity Analysis}
Sensitivity analysis is another commonly used uncertainty quantification approach. By systematically investigating the relationship between input variables and model outputs, sensitivity analysis provides uncertainty quantification in complex systems~\citep{hofer1999sensitivity,cacuci2004comparative} and can be categorized into two classes:
\begin{itemize}
  \item  \textbf{Local sensitivity analysis}: Examines the impact of small changes in parameters on the output~\citep{gustafson1996local}. In other words, it examines how a slight change in one input (while keeping others fixed) influences the output. Typically, this is achieved by computing partial derivatives of the output for each input parameter.\\
 \item   \textbf{Global sensitivity analysis}: Considers the entire input space and assesses the contribution of each input parameter to the output variability~\citep{haaker2004local}. It explores the overall contribution of each input to the variability of the model output. Instead of just looking at small local perturbations, global sensitivity analysis considers the whole range of input uncertainties and how they propagate through the model~\citep{chen2005analytical}.
\end{itemize}

Sampling-based methods play a crucial role in both local and global sensitivity analysis, particularly in global approaches where the focus is on the entire input space. These methods involve generating samples from the input parameter space and then propagating them through the model to observe their impact on the output~\citep{cacuci2004comparative}.
\end{appendices}


\bibliography{Reference}

\end{document}